\documentclass[twocolumn,10pt,amsmath,amssymb,prc,aps,superscriptaddress]{revtex4-1}
\usepackage{graphicx}
\graphicspath{{figs/}}
\usepackage{bm}
\usepackage[T1]{fontenc}
\usepackage{hyperref}
\hypersetup{
 colorlinks  = true, 
 urlcolor   = black, 
 linkcolor  = blue, 
 citecolor  = red  
} 
\usepackage[dvipsnames]{xcolor} 

\usepackage[normalem]{ulem}

\newcommand{\MeV}{\textrm{MeV}}
\newcommand{\fm}{\textrm{fm}}
\newcommand{\kF}{k_\mathrm{F}}
\newcommand{\eF}{\varepsilon_\mathrm{F}}
\newcommand{\minigap}{E_{\mathrm{mg}}}
\newcommand{\kB}{k_\mathrm{B}}
\newcommand{\Tc}{T_\mathrm{crit}}
\newcommand{\WUT}{\mbox{Faculty of Physics, Warsaw University of Technology, Ulica Koszykowa 75, 00-662 Warsaw, Poland}}
\newcommand{\UW}{\mbox{Department of Physics, University of Washington, Seattle, WA 981951560, USA}}
\newcommand{\ULB}{\mbox{Institut d'Astronomie et d'Astrophysique, CP-226, Universit\'e Libre de Bruxelles, 1050 Brussels, Belgium}}

\begin{document} 
\title{Properties of a quantum vortex in neutron matter at finite temperatures}

\author{Daniel P{\k e}cak} \email{daniel.pecak@pw.edu.pl}
\affiliation{\WUT} 
\author{Nicolas Chamel} \email{nicolas.chamel@ulb.be } \affiliation{\ULB}
\author{Piotr Magierski} \email{piotrm@uw.edu} \affiliation{\WUT} \affiliation{\UW}  
\author{Gabriel Wlaz\l{}owski} \email{gabriel.wlazlowski@pw.edu.pl} \affiliation{\WUT} \affiliation{\UW} 

\date{\today}
\begin{abstract}
We have studied systematically microscopic  properties of a quantum vortex in neutron matter at finite temperatures and densities corresponding to different layers of the inner crust of a neutron star. To this end and in preparation of future simulations of the vortex dynamics, we have carried out fully self-consistent 3D Hartree-Fock-Bogoliubov calculations, using one of the latest nuclear energy-density functionals from the Brussels-Montreal family, which has  been developed specifically for applications to neutron superfluidity in neutron-star crusts. 
By analyzing the flow around the vortex, we have determined the effective radius relevant for the vortex filament model. We have also calculated the specific heat in the presence of the quantum vortex and have shown that it is substantially larger than for a uniform system at low temperatures. 
The low temperature limit of the specific heat has been identified as being determined by Andreev states inside the vortex core. We have shown that the specific heat in this limit does not scale linearly with temperature.
The typical energy scale associated with Andreev states is defined by the minigap, which we have extracted for various neutron-matter densities. Our results suggest that vortices may be spin-polarized in the crust of magnetars. 
Finally, we have obtained a lower bound for the specific heat of a collection of vortices with given surface density, taking into account both the contributions from the vortex core states and from the hydrodynamic flow. 
\end{abstract}
\maketitle

\section{Introduction}
The existence of superfluids in neutron stars was conjectured long before the discovery of those compact objects~\cite{migdal1959} (see, e.g., Ref.~\cite{chamel2017} for a recent  overview). 
The first evidence came from accurate radio timing measurements of the rotational frequency of pulsars, revealing sudden spin-ups (whose duration is still not resolved nowadays) followed by relaxations lasting days to months. Such phenomena occurring on a timescale considerably longer than that usually encountered in nuclear-physics experiments can be naturally explained by neutron superfluidity~\cite{baym1969superfluidity}. The frequency 'glitches' themselves are generally interpreted as macroscopic manifestations of the unpinning of neutron quantized vortices in the crust of neutron stars~\cite{anderson1975,pines1985superfluidity}. 

While there is a scientific consensus that superfluidity plays a key role in the glitch phenomenon, many questions still remain open (see, e.g., Ref.~\cite{haskell2015models} for a recent review). This stems from the fact that unlike terrestrial superfluids such as liquid helium and ultracold atomic gases, whose properties can be measured and even experimentally controlled, the cold dense neutron liquid present inside neutron stars cannot be produced on Earth. Although astrophysical observations can indirectly reveal the properties of nuclear superfluids, their interpretation poses a great challenge for theorists. Indeed, it is expected that the global dynamics of neutron stars is determined by the  dynamics of about $10^{18}$ vortices (whose typical core size extends over a length of no more than about few tens of femtometers), averaged over a scale of the order of one kilometer for which general-relativistic effects come into play~\cite{sourie2017global,gavassino2020}. 

The very large differences in scales suggest a bottom-up approach: the motion of large collections of vortices at mesoscopic scales (large compared to intervortex spacing but small compared to the stellar radius) 
could thus be followed using effective Vortex Filament Models (VFM), which have proved their usefulness for the description of vortices in superfluid ${}^4$He~\cite{schwarz1,schwarz2,schwarz3}. However, one has to remember that there are two important differences between vortices in ${}^4$He and in neutron stars. Contrary to bosonic vortices, fermionic ones have their core filled with matter in a normal state. 
Moreover, fermionic vortices allow for yet another degree of freedom, which plays a crucial role for 
vortex structure. Namely, the spin imbalance, which in the case of ultracold gases is routinely investigated in laboratories~\cite{zwierlein2006imbalance},
will affect both the internal structure of the core and its size~\cite{drummond,magierski_vortex}. 
In neutron stars,  spin polarization can be potentially induced by strong magnetic fields, but
it remains an open question whether it is strong enough to affect the structure of the vortex core.
Although the finite size of vortices can be ignored at mesoscopic scales, their quantum structure (as effectively embedded in model parameters) may still have important implications for the stellar dynamics~\cite{feibelman1971}. A better understanding of the microscopic physics underlying effective VFM is therefore highly desirable. 

At the smallest scales of interest, which are related
to the vortex core size and intervortex spacing, the motion of individual vortices is nonrelativistic (their velocity is small compared to  the speed of light~\cite{gugercinoglu2016} and space-time curvature is negligible~\cite{glendenning1997}) so that the methods developed in condensed-matter physics to study laboratory superfluids can be directly adapted to dense stellar environments. 
The microscopic structure of a single neutron vortex was previously studied at  zero  temperature solving the Bogoliubov-de Gennes equations~\cite{de1999microscopic} (i.e. the single-particle Hamiltonian is that of noninteracting particles)
with an effective density-dependent contact pairing interaction fitted to many-body calculations in homogeneous neutron matter using bare nucleon-nucleon potentials. More realistic types of calculations involved the nuclear-energy density functional (EDF) theory within the Hartree-Fock-Bogoliubov (HFB) method~\cite{yu2003} using the FaNDF$^0$  functional~\cite{fayans1998towards,fayans2000nuclear} supplemented with a pairing functional adjusted to $^1$S$_0$ pairing gaps in homogeneous neutron-matter. However, these calculations focused on very dilute neutron matter at densities corresponding to the shallow layers of the inner crust of neutron stars. The pinning of a vortex by a nuclear cluster was later studied in Ref.~\cite{avogadro2007} by solving the HFB equations in a cylindrical cell using the Skyrme SII functional~\cite{SkyrmeSII} for the normal part, and the phenomenological parametrization of Ref.~\cite{garrido1999effective} for the pairing part. More recently, studies of the dynamics of a vortex have been undertaken by solving the fully three-dimensional and symmetry-unrestricted time-dependent HFB equations~\cite{wlazlowski2016vortex}. 

As a first step towards realistic simulations of the vortex dynamics in neutron-star crusts, we present in this paper finite-temperature HFB calculations of a single vortex in an otherwise homogeneous neutron superfluid using the Brussels-Montreal EDF BSk31~\cite{chamel2016further}, which was specifically constructed for astrophysical applications. The contributions from time-odd mean fields, which may play an important role in the superfluid dynamics~\cite{allard2019,allard2020entrainment}, are taken into account for the first time. 

The paper is organized as follows: in Sec.~\ref{sec:system} we describe the basic ingredients
of the approach and the techniques used together with our numerical setup. Then we describe our main results: in Sec.~\ref{sec:length} we describe the length scales associated with the vortex at different temperatures, subsequently the discussion of velocity fields and superfluid fraction is included in Sec.~\ref{sec:velocity}. Finally, in Sec.~\ref{sec:heat_capacity} the specific heat of neutron matter for densities corresponding to the inner crust in the presence of the vortex is discussed. 
We summarize this article in Sec.~\ref{sec:conclusions}.

\section{Hartree-Fock-Bogoliubov equations and numerical setup 
}\label{sec:system}

\subsection{Nuclear energy-density functional theory}

The EDF theory has been successfully applied for the description of both the 
static and dynamic properties of finite nuclei, as well as for nuclear reactions \cite{bender2003self, nakatsukasa2016, bulgac2016induced, magierski2019nuclear}. 

The building blocks of the semi-local energy-density functional $\mathcal{E}$ we consider here, comprise of the following local densities and currents (for each nucleon species): particle number density $\rho(\pmb{r})$, kinetic density $\tau(\pmb{r})$, anomalous density $\nu(\pmb{r})$, and momentum density $\pmb{j}(\pmb{r})$, all of which are defined through the Bogoliubov quasiparticle amplitudes $u_{n,\sigma}(\pmb{r}), v_{n,\sigma}(\pmb{r})$, where $n$ represents the relevant set of quantum numbers while $\sigma=\{\uparrow,\downarrow\}$ denotes the spin components. The spin-orbit coupling, which plays a minor role in neutron-star crusts (see, e.g., the discussion in Ref.~\cite{pearson2018}), will be omitted hereafter. 
Consequently, the quasiparticle amplitudes corresponding to spin-up and spin-down components are related to each other and one needs
to calculate $u_{n,\uparrow}(\pmb{r}), v_{n,\downarrow}(\pmb{r})$ amplitudes only.
The densities and currents read (see Refs. \cite{bulgac2012, magierski2019nuclear}
for details)

\begin{align}\label{eq:densities}
 \rho(\pmb{r})  &= 2\sum_{n} \left[ |v_{n,\downarrow}(\pmb{r})|^2 f_T(-E_n) 
                                + |u_{n,\uparrow}(\pmb{r})|^2 f_T(E_n) \right],         \\
 \tau(\pmb{r})  &= 2\sum_{n} \left[ |\pmb{\nabla} v_{n,\downarrow}(\pmb{r})|^2 f_T(-E_n)
                                + |\pmb{\nabla} u_{n,\uparrow}(\pmb{r})|^2 f_T(E_n) \right],  \\
 \nu(\pmb{r})   &= 2\sum_{n} u_{n,\uparrow}(\pmb{r}) v^*_{n,\downarrow}(\pmb{r})
                 ( f_T(-E_n) -  f_T(E_n) ),     \\
 \pmb{j}(\pmb{r}) &= 2\sum_{n} \mathrm{Im} \left[ v_{n,\downarrow}(\pmb{r}) \pmb{\nabla}  
                 v^*_{n,\downarrow}(\pmb{r}) \right ] f_T(-E_n) + \nonumber \\
     &+ \sum_{n}\mathrm{Im} \left[ u_{n,\uparrow}(\pmb{r}) \pmb{\nabla} u^*_{n,\uparrow}(\pmb{r}) 
              \right ] f_T(E_n)  ,     
\end{align}
where the summations are performed over positive quasiparticle energies with a suitable  regularization to avoid divergences, as will be discussed in Section~\ref{subsec:cutoff}. 
In order to allow for the description at finite temperatures, the following thermal occupation factors have been introduced:
\begin{align}
f_T(E)=\left[1+\exp\left(\frac{E}{T}\right)\right]^{-1} ,
\end{align}
(setting Boltzmann's constant $\kB=1$). 
The quasiparticle amplitudes and quasiparticle energies $E_{n}$
fulfill the HFB equations, which in the coordinate-space representation take the following form for all temperatures:
\begin{equation}\label{eq:hfb}
\left(
\begin{array}{cc}
 h(\pmb{r})-\mu & \Delta(\pmb{r})  \\  \Delta^*(\pmb{r}) & -h^*(\pmb{r})+\mu  
\end{array}
\right)
\left(
\begin{array}{c}
 u_{n,\uparrow}(\pmb{r}) \\  v_{n,\downarrow}(\pmb{r}) 
\end{array}
\right)
=E_n
\left(
\begin{array}{c}  
u_{n,\uparrow}(\pmb{r}) \\  v_{n,\downarrow}(\pmb{r}) 
\end{array}
\right),
\end{equation}
where $\mu$ is the chemical potential. 
The single-particle fields are defined via the variational principle:
\begin{align}
 h(\pmb{r}) &= \frac{\delta\mathcal{E}}{\delta\rho} 
 - \pmb{\nabla} \frac{\delta\mathcal{E}}{\delta\tau} \cdot \pmb{\nabla} 
 - \frac{i}{2} \left\lbrace \frac{\delta\mathcal{E}}{\delta {\pmb{j}}},\pmb{\nabla} \right\rbrace,
\\
 \Delta(\pmb{r}) &= -2\frac{\delta\mathcal{E}}{\delta\nu^*} ,
\end{align}
where $\{.,.\}$ denotes the anticommutator, and $\delta\mathcal{E}/\delta {\pmb{j}}$ means a vector constructed by variation over three components of the current $\pmb{j}$. 
The contribution of the mean potential vector field induced by currents to the single-particle Hamiltonian $h(\pmb{r})$ 
was ignored in previous calculations.

The EDF is the central element of the theory and encodes information about nuclear interactions. It is expressed through the densities and currents ($\rho, \tau, \nu, \pmb{j}$) and has the generic form:
\begin{align}\label{eq:density_functional}
 \mathcal{E}(\rho,\pmb{\nabla}\rho,\nu,\tau,{\pmb{j}})
 &= \frac{\hbar^2}{2 M} \tau 
 + \mathcal{E}_\rho(\rho) + \mathcal{E}_{\Delta\rho}(\rho,\pmb{\nabla}\rho) \\
 & + \mathcal{E}_\tau(\rho,\tau,{\pmb{j}}) \nonumber
 + \mathcal{E}_\pi(\rho,\pmb{\nabla}\rho,\nu).
\end{align}
The first term corresponds to the kinetic energy density for a nucleon with mass $M$, while consecutive terms represent the energy density due to nuclear interactions. The energy densities $\mathcal{E}_\rho$ and $\mathcal{E}_{\Delta\rho}$ are related to the interaction of the nucleons with the background density  and its fluctuations respectively. The next term $\mathcal{E}_\tau$ is associated with a density-dependent effective mass and also gives rise to current-current couplings (so called entrainment effects)~\cite{allard2020entrainment}. The last term describes the pairing energy density $\mathcal{E}_\pi$. The contribution related to the spin-orbit coupling has been omitted in
the presented calculations. This term has been shown to play a minor role in the context of stellar 
environment where spatial density fluctuations are much smaller than 
in the case of finite nuclei~\cite{pearson2015,pearson2018,mondal2020}.

\subsection{Brussels-Montreal nuclear-energy density functionals}
\begin{figure}[!ht]
\centering
\includegraphics[width=0.99\linewidth]{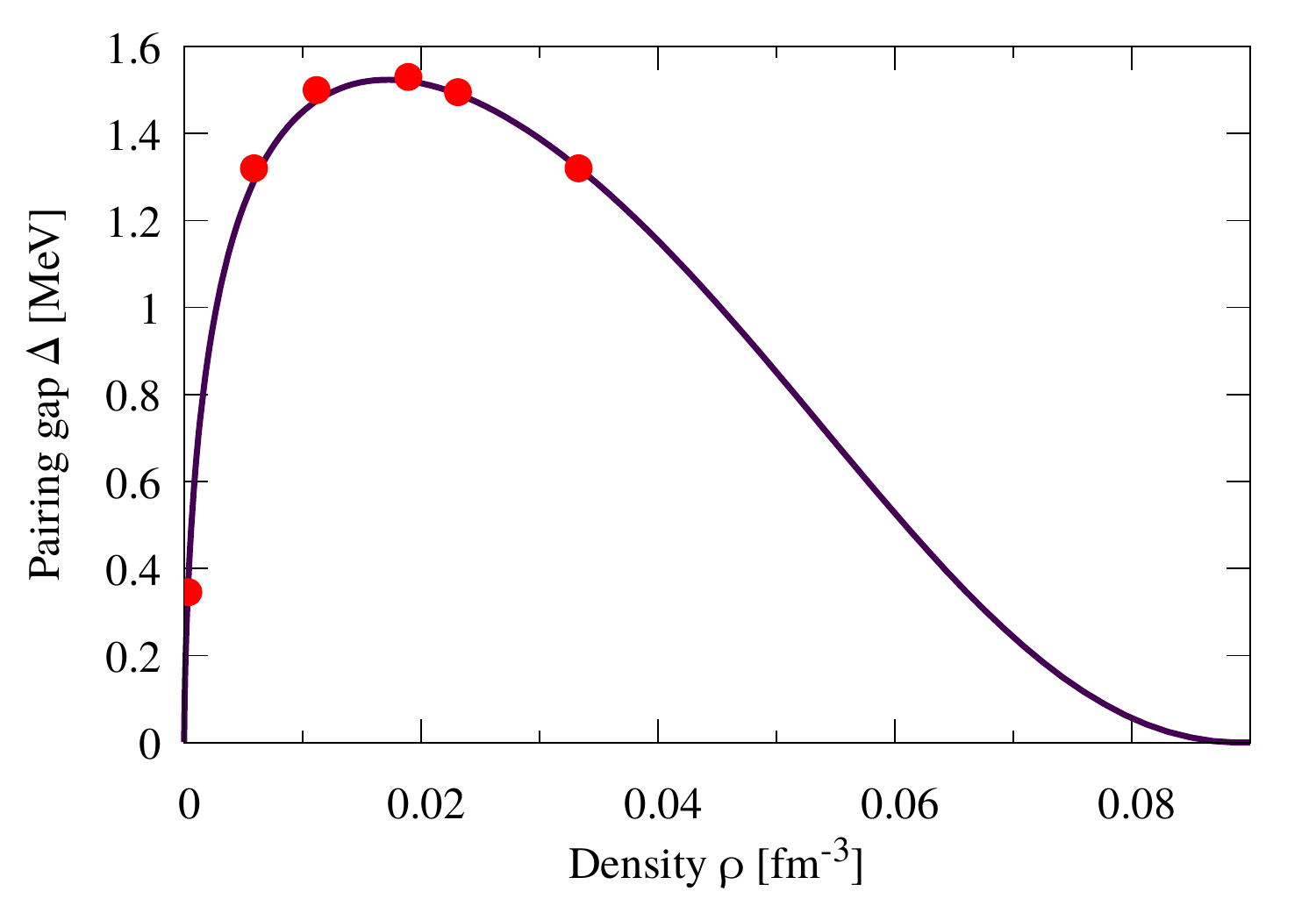}
\caption{
$^1$S$_0$ pairing gaps in uniform neutron matter as a function of density $\rho$, as calculated in Ref.~\cite{cao2006screening} including both medium polarization effects and self-energy corrections. The six red points indicate the densities considered in this article.
\label{fig:delta}}
\end{figure}

The Brussels-Montreal functionals, based on generalized Skyrme effective interactions, were 
not only precision-fitted to experimental nuclear 
data (e.g. binding energies, radii), but were specifically constructed for applications 
to extreme astrophysical environments. In particular, properties of uniform neutron matter, 
as determined from many-body calculations using bare nucleon-nucleon potentials, were 
included in their fit. Some of these functionals have been already employed to calculate 
in a unified and thermodynamically consistent way the internal constitution of neutron stars 
and their equation of state, from the surface to the core~\cite{fantina2013,pearson2018}. Except 
for BSk19, most recent Brussels-Montreal functionals have been shown to be compatible with existing astrophysical observations, including the latest constraints inferred from analyses of the 
gravitational-wave signal GW170817~\cite{perot2019}. For our present purpose, we have 
adopted the functional BSk31~\cite{chamel2016further}. This functional was not only accurately 
fitted to the $2353$ measured atomic masses of nuclei with proton and neutron numbers $\geq 8$  (taken from the 2012 Atomic Mass Evaluation \cite{Wang}) with a root-mean-square deviation of about $0.6$ MeV, but it was also adjusted to microscopic calculations of the equation of state, effective masses and most importantly $^1$S$_0$ pairing gaps  of pure neutron matter. Therefore, this 
functional appears to be particularly well-suited for the study of neutron superfluidity in  neutron-star crusts. The pairing gaps of Ref.~\cite{cao2006screening}, to which the BSk31 functional was fitted, are depicted in Fig.~\ref{fig:delta} and were obtained from diagrammatic calculations taking into account medium-polarization and self-energy effects. The figure also indicates the  densities that will be considered in this paper. These representative densities span different layers of the crust of neutron stars. 

Focusing on neutron matter, the form of the respective contributions to the total energy in Eq.~\eqref{eq:density_functional} for the BSk31 functional (see Ref.~\cite{chamel2016further} for details) is the following:
\begin{align}
\mathcal{E}_\rho(\rho) 
&= C^\rho \rho^2
\\
\mathcal{E}_{\Delta\rho}(\rho,\pmb{\nabla}\rho) 
&= - \left(\pmb{\nabla}\rho\right)^2 
  C^{\Delta\rho} \\
\mathcal{E}_\tau(\rho,\tau,{\pmb{j}}) 
&= C^\tau (\rho\tau - {\pmb{j}}^2)
\\
\mathcal{E}_\pi(\rho,\pmb{\nabla}\rho,\nu)
&=\frac{1}{4}  v^{\pi}(\rho)   \nu\nu^{*}
 + \kappa|\nabla\rho|^2,
\end{align}
where the term $\kappa|\nabla\rho|^2$ (see Eq.~(9) in Ref~\cite{chamel2009pairing}) in the pairing energy density will be neglected hereafter. 
The coupling coefficients $C^\rho,C^\tau,C^{\Delta\rho}$ depend on density~$\rho$, and are defined in Appendix~\ref{appendix}, see Eqs.~(\ref{eq:a:crho})-(\ref{eq:a:claplace}).
The pairing strength is defined according to Ref.~\cite{chamel2009pairing}:
\begin{equation}
v^{\pi}(\rho) = - \frac{8 \pi^2}{I} \left(\frac{\hbar^2}{2 M^\oplus}\right)^{3/2},
\label{eqn:vpi}
\end{equation}
where the effective mass $M^\oplus$ is defined via Eq.~\eqref{eq:effective_mass}. We use the following  approximate analytic formula of Ref.~\cite{chamel2010effective} for the denominator:
\begin{equation}
I = \sqrt{\mu} \left[ 2 \ln{\left( \frac{2 \mu}{|\Delta|} \right)} + \Lambda\left( \frac{\varepsilon_\Lambda}{\mu} \right) \right],
\end{equation}
where $\varepsilon_\Lambda$ is a cutoff that must be introduced to avoid ultraviolet divergences, 
as will be discussed in the next section. 
The function $\Lambda(x)$ reads:
\begin{equation}
\Lambda(x) = \ln(16x) + 2\sqrt{1+x} - 2 \ln\left({1+\sqrt{1+x}}\right) - 4.
\label{eqn:Lambda}
\end{equation}

The advantage of the present formulation as compared to previous calculations using phenomenological parametrizations such as that of Garrido et al.~\cite{garrido1999effective} fitted to a specific value of the cutoff is that the pairing strength adopted here is automatically renormalised for different choices of the cutoff.

\subsection{Convergence and cutoff}
\label{subsec:cutoff}
The regularization scheme associated with the local pairing functional, defined via Eqs~(\ref{eqn:vpi})-(\ref{eqn:Lambda}), was originally constructed~\cite{chamel2008} by cutting off single-particle energies lying above some limiting value $\varepsilon_\Lambda$ (as measured with respect to the chemical potential),  which was set to 6.5 MeV for the functional BSk31~\cite{chamel2016further}. 
Although such a low value allows for fast static HFB calculations, it is not suitable for reliable time-dependent HFB simulations due to the violation of energy conservation~\cite{magierski2019nuclear}. In anticipation of future studies, we shall adopt here values of order $\varepsilon_\Lambda\sim 100$~MeV. In this case, we thus have $\varepsilon_\Lambda\gg |\Delta|$ therefore to a good approximation we can impose the same cutoff on the quasiparticle energies directly, i.e. by summing over quasiparticle states such that $E_n< E_c\approx \varepsilon_\Lambda$. 

\begin{figure}[!ht]
\centering
\includegraphics[width=0.99\linewidth]{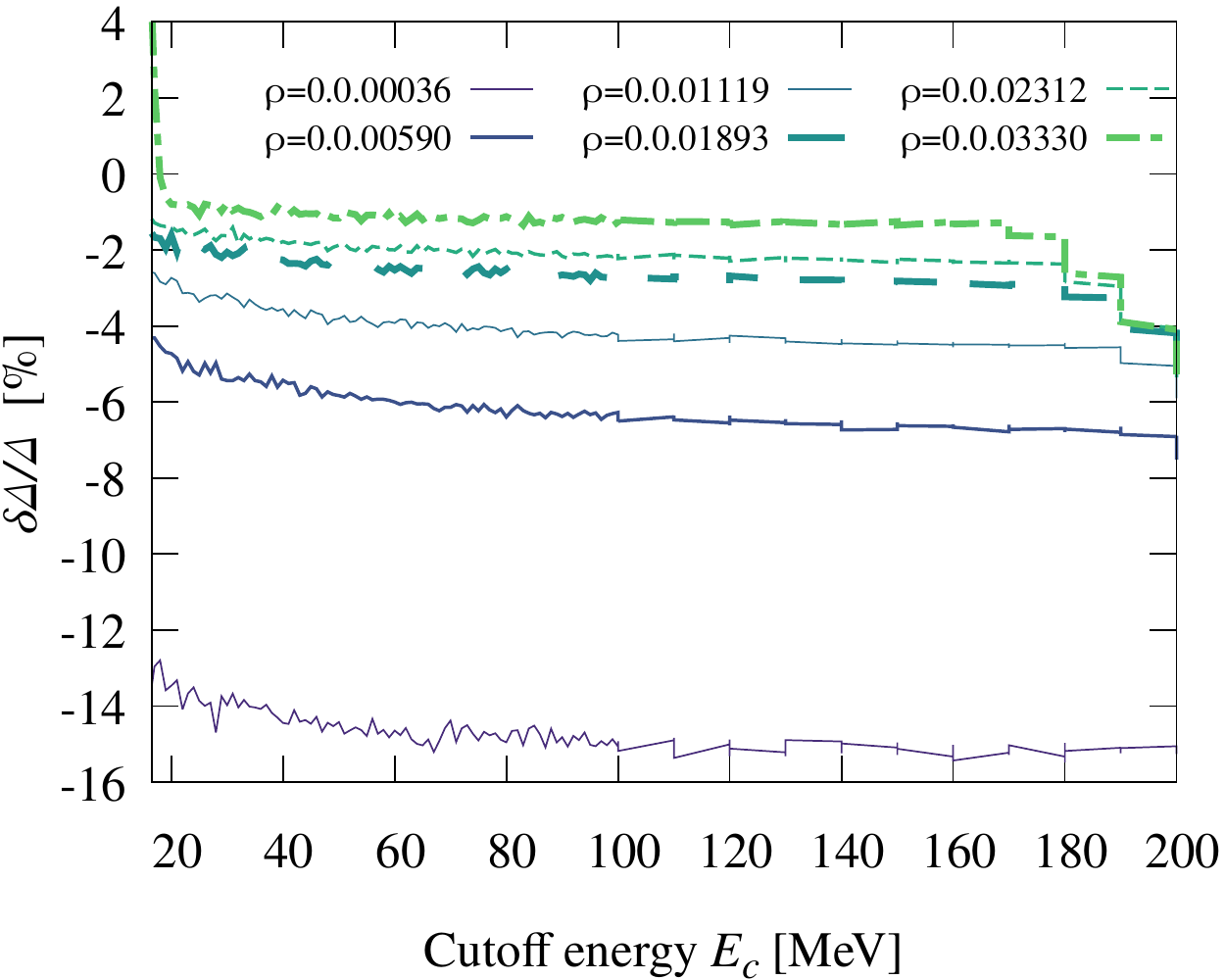}
\caption{
Relative deviations $\delta\Delta/\Delta$ between the calculated $^1$S$_0$ pairing gap in uniform neutron matter and the reference one as a function of the cutoff energy $E_c$ for different densities $\rho$ in fm$^{-3}$.  
As expected from the weak-coupling approximation $|\Delta| \ll \mu$ underlying the adopted pairing functional, the error is the highest for the lowest density. 
In the vicinity of the maximum pairing gap, the overall precision of the self-consistent calculations lies within $5\%$.
\label{fig:cutoff}}
\end{figure}

Let us  also recall that the analytical pairing strength, Eqs~(\ref{eqn:vpi})-(\ref{eqn:Lambda}), was obtained under the assumptions that $|\Delta|\ll \mu$ and $\varepsilon_\Lambda \gg |\Delta|$. 
To assess the reliability of these approximations and the validity of our computer code, we have  recalculated the pairing gaps $\Delta$ in uniform neutron matter for different cutoff energies $\varepsilon_\Lambda$ (approximated by $E_c$) and we have compared our results with the reference pairing gaps of Ref.~\cite{cao2006screening} shown in  Fig~\ref{fig:delta}. 
We have solved the HFB equations in a cubic box with periodic boundary conditions on a Cartesian grid of $128^3$ points with a spacing $\delta x=\delta y=\delta z=1$~fm. 
Results are shown in Fig.~\ref{fig:cutoff} for different densities. 
With increasing cutoff starting from $E_c=16.5$~MeV, one can see that the error in the pairing gap varies rapidly within $2-15\%$ depending on the density but remains fairly independent of the cutoff above $E_c\approx 40$ MeV.
The sudden deterioration of the precision around $E_c=180$ MeV arises from the discretization of space. Indeed, a finite grid spacing $\delta x$ prohibits wave numbers higher than $k_c=\pi/\delta x$, which translates into an energy cutoff $E_c\approx \hbar^2 k_c^2/(2 M)\approx 200$~MeV.

\subsection{Numerical setup}\label{num}
Our numerical setup to simulate a vortex is similar to that previously adopted in Ref.~\cite{wlazlowski2016vortex}. Namely, we introduce an axially symmetric external 
potential to confine the neutron superfluid in a tube. 
The potential is parametrized as follows (distances are given 
in $\fm$, and energies in $\MeV$): 
\begin{equation}
V_{\mathrm{ext}}(r) = \left\{
  \begin{array}{ll}
    0 & \mbox{if } \phantom{ r_1 < } r < r_1, \\
    50 s(r-r_1,r_2-r_1)   & \mbox{if } r_1 < r < r_2, \\
    50 & \mbox{if } r_2 < r \phantom{r_1 <},
  \end{array}
\right.
\end{equation}
where by $r$ we denote the distance from the $z$ axis.
The smooth switching function $s$ in the transition zone is defined as follows:
\begin{equation}
s(r,R) = \frac{1}{2} \left\{1 + \tanh\left[\tan\left( \pi \frac{r}{R} - \frac{\pi}{2}\right) \right]  \right\}.
\end{equation}
In Ref.~\cite{wlazlowski2016vortex}, the volume of the box was chosen to be $V=60\times 60\times 75$ fm$^3$, 
and the parameters of the external potential were set to $r_1=30$~fm, $r_2=35$~fm. To reduce as much as possible the influence of this potential on the superfluid properties, we consider 
here a cubic box with a larger volume $V=(90$ fm$)^3$. For our setup, $r_1$ and $r_2$ are now $40$ and $44\,\fm$ respectively. 
The spatial resolution is set to $1.5$ fm in each direction, as in Ref.~\cite{wlazlowski2016vortex}. 
This corresponds to a maximum accessible momentum $\hbar k_c \approx 400$~MeV/c, or equivalently, a cutoff energy $E_c\approx 90$ MeV.

\begin{figure}[ht]
\centering
\includegraphics[width=0.99\linewidth,trim=0mm 0 0 0,clip]{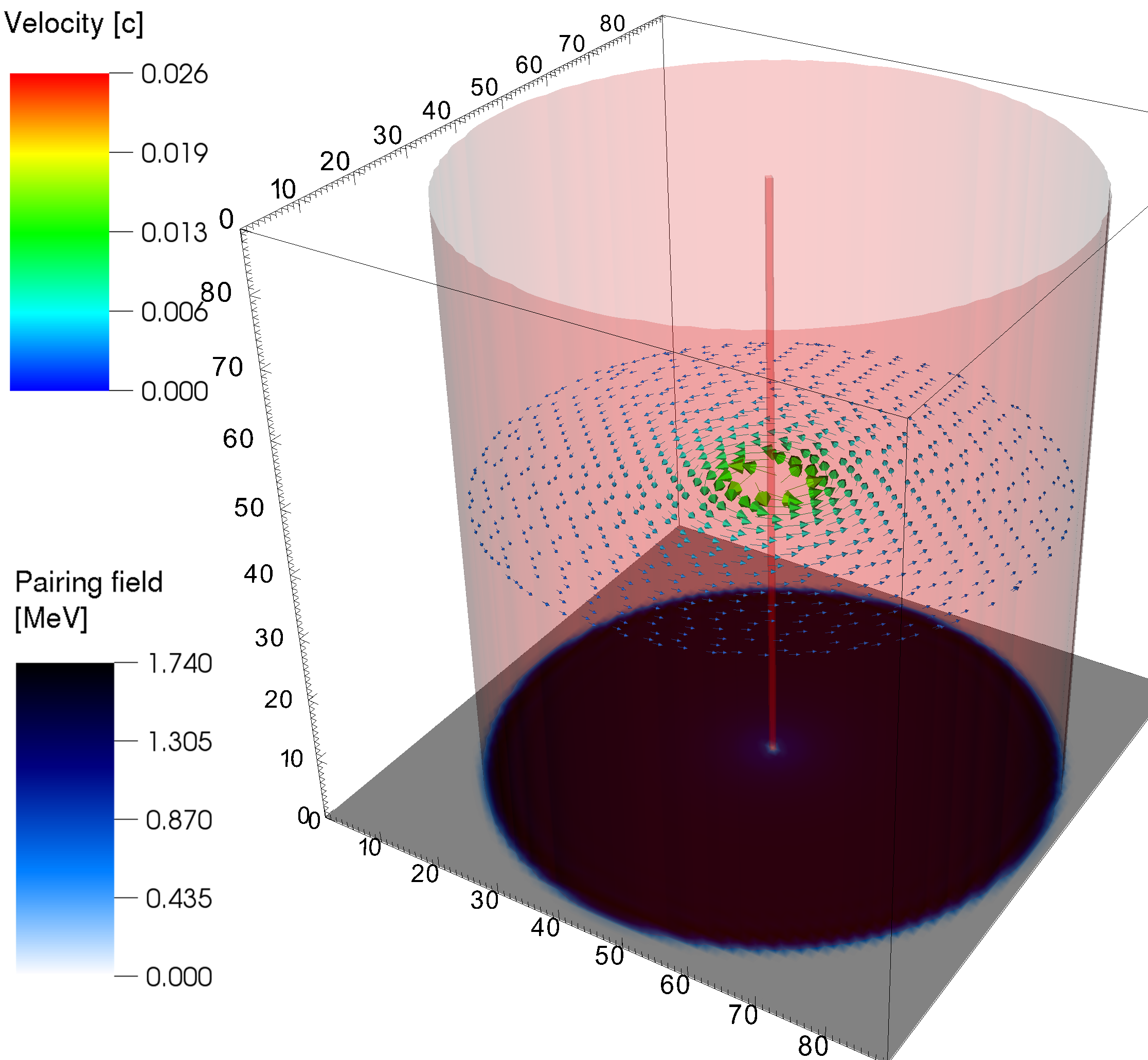}
\caption{
Three-dimensional view of a vortex in neutron matter at density $\rho=0.0189\mathrm{ fm}^{-3}$. The red translucent contour marks the region where the modulus of the pairing field $|\Delta(\pmb{r})|$ corresponds to 15\% of the bulk pairing gap $\Delta_{\infty}=1.55 \mathrm{MeV}$. The line at the center shows the vortex core, while the outer tube is connected to boundary effects and roughly delimits the range of the external potential.
Value of $|\Delta(\pmb{r})|$ are plotted at the bottom of the cube (the system is translationally invariant along the $z$ axis). Arrows at mid-height show the vector current $\pmb{j}(\pmb{r})$.
\label{fig:visit}}
\end{figure}

A vortex is generated at the center of the tube by imposing 
constraints on the phase of the pairing field $\Delta(\pmb{r})=|\Delta(r)|e^{i\phi}$, where $\phi=\arctan(y/x)$ is the azimuthal angle. Although we consider here a single vortex (with a winding number equal to 1), our computer code is flexible enough to allow for multiple vortices with higher winding numbers at no additional cost. 
Due to the symmetry along the $z$ direction, the system is effectively two-dimensional. In Fig.~\ref{fig:visit} we show red translucent contours corresponding to $\approx 0.19 \mathrm{MeV}$ of the pairing gap. 
Due to the presence of the vortex and boundaries, the pairing field is nonuniform and vanishes in
the  vortex core and at the boundaries of the tube. Therefore, the thin red line in Fig.~\ref{fig:visit} can be associated with the vortex line, while the red tube roughly corresponds to the external potential. 
The amplitude of the pairing field $|\Delta(\pmb{r})|$ is shown at the bottom of the tube. 

In the calculations of the specific heat we changed the geometry described above. Namely, we have considered
the volume of $V=180 \times 180 \times 36$ fm$^3$, for which we set the external potential parameters accordingly higher to $r_1=80$~fm, $r_2=87.5$~fm.
These modifications were introduced in order to minimize the influence of boundary effects and to make
sure that the density oscillations induced by the vortex are not influenced by the boundary conditions.

\section{Length scales}\label{sec:length}


Fermionic systems are characterized by two different characteristic length scales. 
One is connected to the inverse of the Fermi momentum $\kF^{-1}=(3\pi^2\rho)^{-1/3}$. The other is specific to superfluid systems: the coherence length $\xi=\hbar^2 \kF/(M \pi |\Delta|)$. 
In the context of neutron-star crusts, these two length scales are well separated. Even for the largest
pairing gap predicted in dilute neutron matter, the coherence length is a few times larger than the average interparticle distance measured by $\kF^{-1}$.
The only system known in Nature, where these two scales become comparable is the unitary Fermi gas realized in ultracold atomic systems~\cite{BECBCS}. 
Nevertheless, dilute neutron matter in neutron-star crusts is the nuclear system that shares many similarities with the unitary Fermi gas. 
Separation of length scales has two important consequences for the vortex structure. First, the spatial length scales over which the pairing field changes will be much larger than those related to the normal density fluctuations. Second,
the density of Andreev states inside the vortex core will be significant
\cite{Sensarma2006, magierski_vortex,elgaroy2001superfluid}. This has important implications for the calculation 
of the specific heat~\cite{caroli1964}, as will be shown explicitly in Section~\ref{sec:heat_capacity}. 

\begin{figure}
\centering
\includegraphics[width=0.99\linewidth]{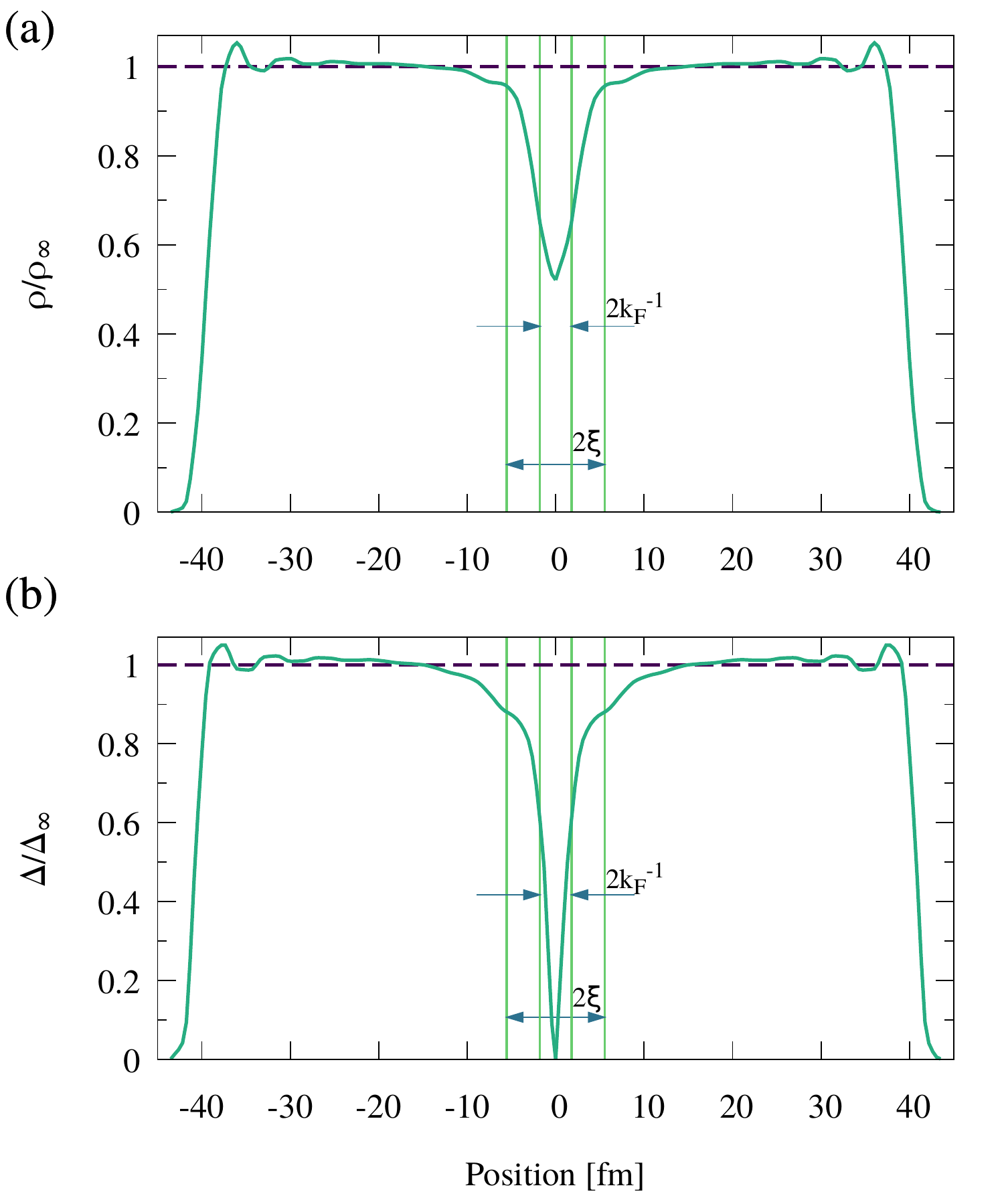}
\caption{
Section through the center of a vortex core inside neutron matter. The upper figure shows the density profile (a), and the lower figure the pairing field (b). The neutron density 
and the modulus of the pairing field far from the vortex are $\rho_\infty=0.0059$ fm$^{-3}$ and $\Delta_\infty=1.334$ MeV, respectively. The two length scales of the system are depicted in the plots: the coherence length $\xi$ and the inverse of the Fermi wave vector $\kF^{-1}$.
\label{fig:core}}
\end{figure}

The top and bottom  panels of  Fig.~\ref{fig:core} display the density profile of a vortex and the associated pairing field, respectively. Characteristic length scales are indicated by vertical lines.
In bosonic superfluids, the order parameter is directly related to the density, and as a consequence both of them vanish in the vortex core. This is not the case for fermionic superfluids: although the pairing field, which is related to the order parameter (see, e.g., Eq.(29) of Ref.~\cite{allard2020entrainment}) vanishes, only a partial depletion of the density is observed in the vortex core, as previously shown for a neutron vortex in Ref.~\cite{yu2003}.
The existence of a finite density can be traced back to the
occupation of Andreev states. Namely, all Andreev states have a nonzero component of angular momentum along the $z$-axis (vortex line), 
and consequently the density distribution associated with these states exhibits a minimum in the core. The situation changes 
when one allows for spin polarization, as could be induced by a strong magnetic field in magnetars~\cite{PhysRevC.93.015802}. In this case, the majority of the spin particles start to occupy 
states with reversed angular momenta including the state with $0\hbar$.
Consequently, the density minimum in the vortex core immediately disappears~\cite{magierski_vortex}. 

As can be seen in Fig.~\ref{fig:core}, the density variations within the vortex core 
are relatively small, thus justifying the neglect of the spin-orbit coupling (as in  previous studies). The vanishing of the pairing field at the vortex center suggests that the vortex core is a ``normal'' fluid, as will be explicitly shown in the next section by computing the superfluid fraction. 
It may be noted that the behavior of the pairing field in the core seems to change within two length
scales. Deep inside the core the variation is more rapid, and occurs within a length scale set by $\kF^{-1}$, whereas at larger distances it varies at distance of the order of the coherence length.
The presence of these two length scales in the vicinity of the core 
has been previously  discussed in Ref.~\cite{Sensarma2006} in the context of atomic  gases.
Far enough from the center of the vortex core, the local density and pairing field saturate at values characteristic for a uniform system. This can be clearly seen in Fig.~\ref{fig:core}: within few coherence lengths from the vortex core, the system tends to be homogeneous. 
The oscillations seen near the boundaries of the system are due to the imposed external potential. They produce fluctuations on a scale of the order of $\kF^{-1}$ and behave similarly to Friedel oscillations.

\begin{figure}[!ht]
\centering
\includegraphics[width=0.99\linewidth]{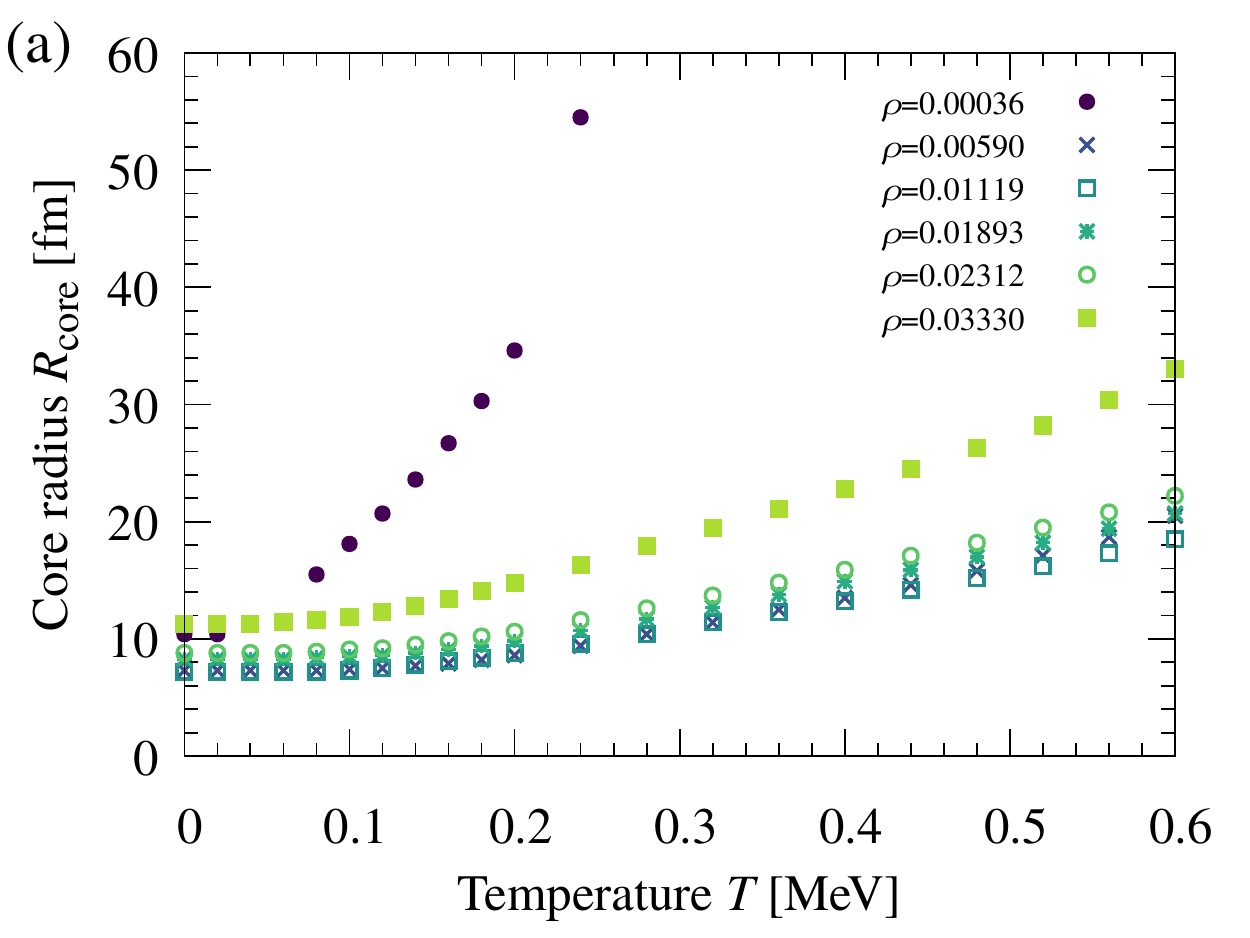}
\includegraphics[width=0.99\linewidth]{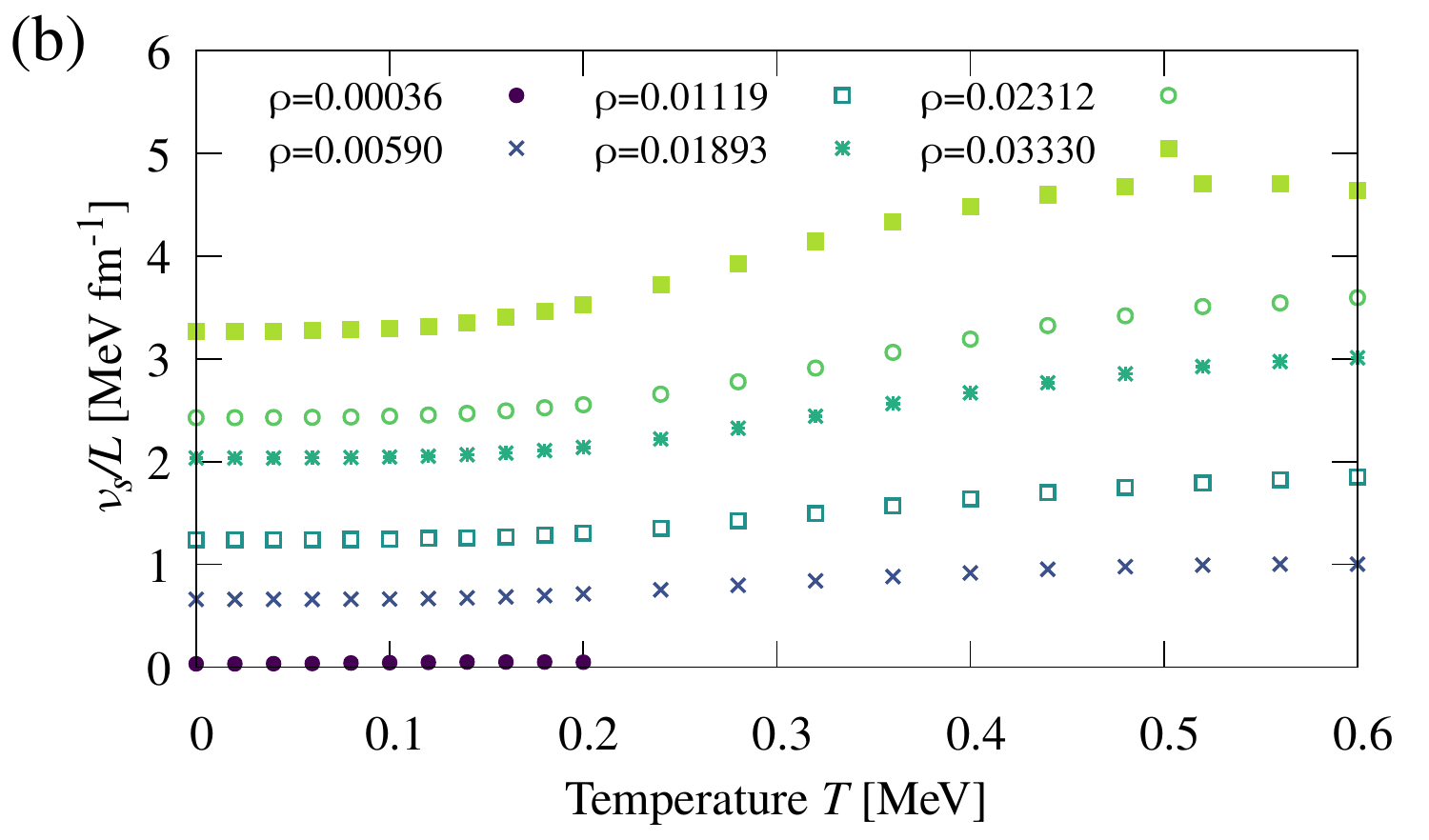}
\caption{
(a) Size of the vortex core $R_\mathrm{core}$ in neutron matter as a function of temperature $T$ for different densities $\rho$. The size of the vortex increases with temperature and diverges at the critical point.
(b) Tension per unit length of the vortex line as a function of temperature $T$ for different densities $\rho$. The tension grows with density and vanishes at the critical point.
\label{fig:coresize}}
\end{figure}

While Fig.~\ref{fig:core} presents zero-temperature results, we have also performed calculations at finite temperatures. To better characterize the structure of a vortex, we have computed the core radius $R_\mathrm{core}$ defined as the distance $r$ at which the pairing field increases to 90\% of the bulk value far from the vortex. The systematic study of $R_\mathrm{core}$ as a function of temperature $T$ and density $\rho$ is summarized in  Fig.~\ref{fig:coresize}(a). Our zero-temperature results are of the same order as those obtained in previous studies~\cite{de1999microscopic,avogadro2008vortex}. In general, thermal effects lead to a reduction of the pairing field thus resulting in a larger vortex core, whose size eventually diverges at the critical temperature $\Tc = |\Delta|/1.76$ (the vortex disappearing in that case). In the finite system we are considering, this means that $R_\mathrm{core}$ increases beyond the size of the box. 

We have also carried out systematic calculations of the vortex tension per unit length. The tension is defined as the difference of energies between the vortex and uniform matter for a given density and temperature. The results are plotted in Fig.~\ref{fig:coresize}(b). The tension of a vortex line is found to increase with density and temperature.

\section{Velocity field and Superfluid Fraction}\label{sec:velocity}

At a distance of a few coherence lengths from the vortex, both the density and absolute
value of the pairing field becomes essentially uniform.
Still, the presence of superflow makes the critical difference between the 
system with and without the vortex even at large distances from the core. 
The presence of the superflow can be characterized by
the velocity field which decays like $1/r$ and is a consequence of a particular phase
pattern of the pairing field. The presence of the superflow is responsible for long range
interaction between vortices and, as can be seen in the next section, produces also a correction
to the specific heat of the system.

\begin{figure}
\centering
 \includegraphics[width=\linewidth]{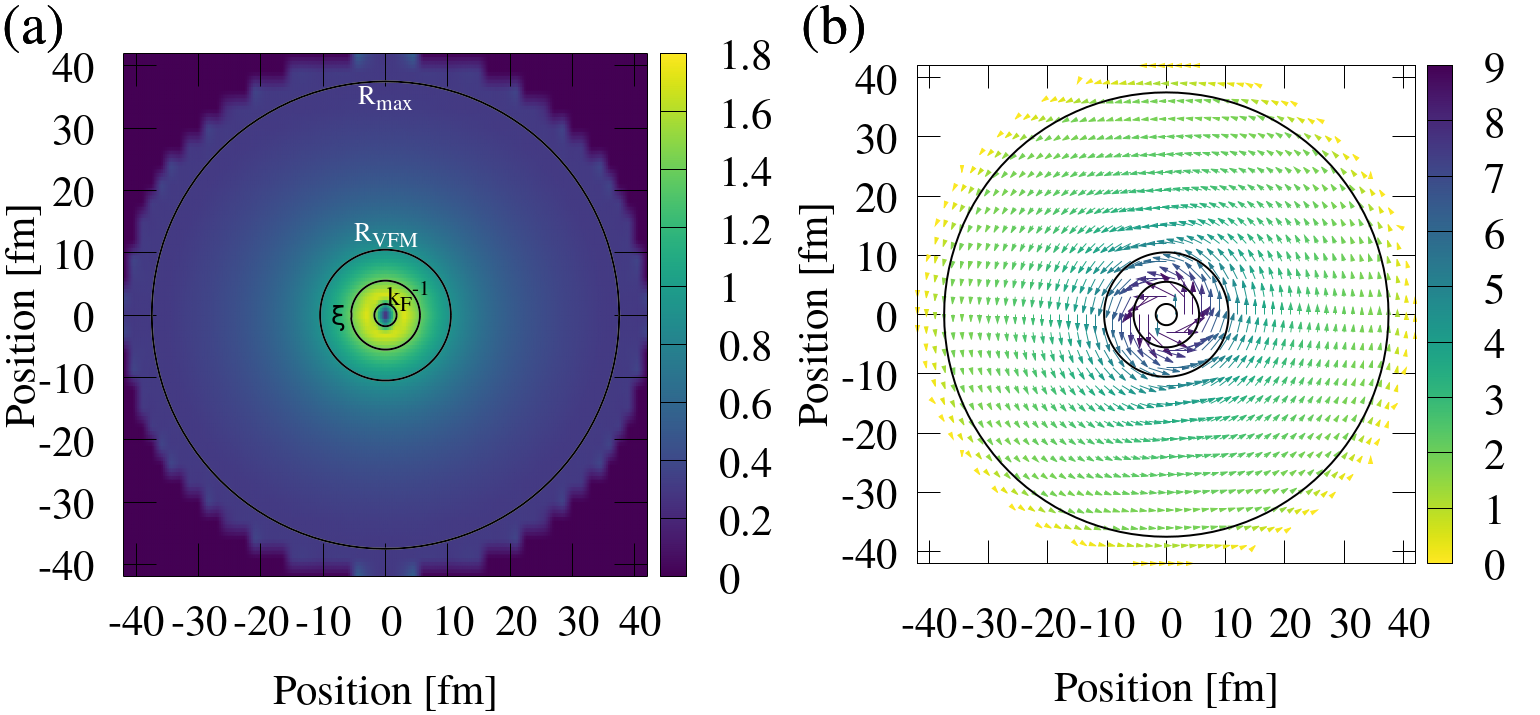}\\
 \includegraphics[width=\linewidth]{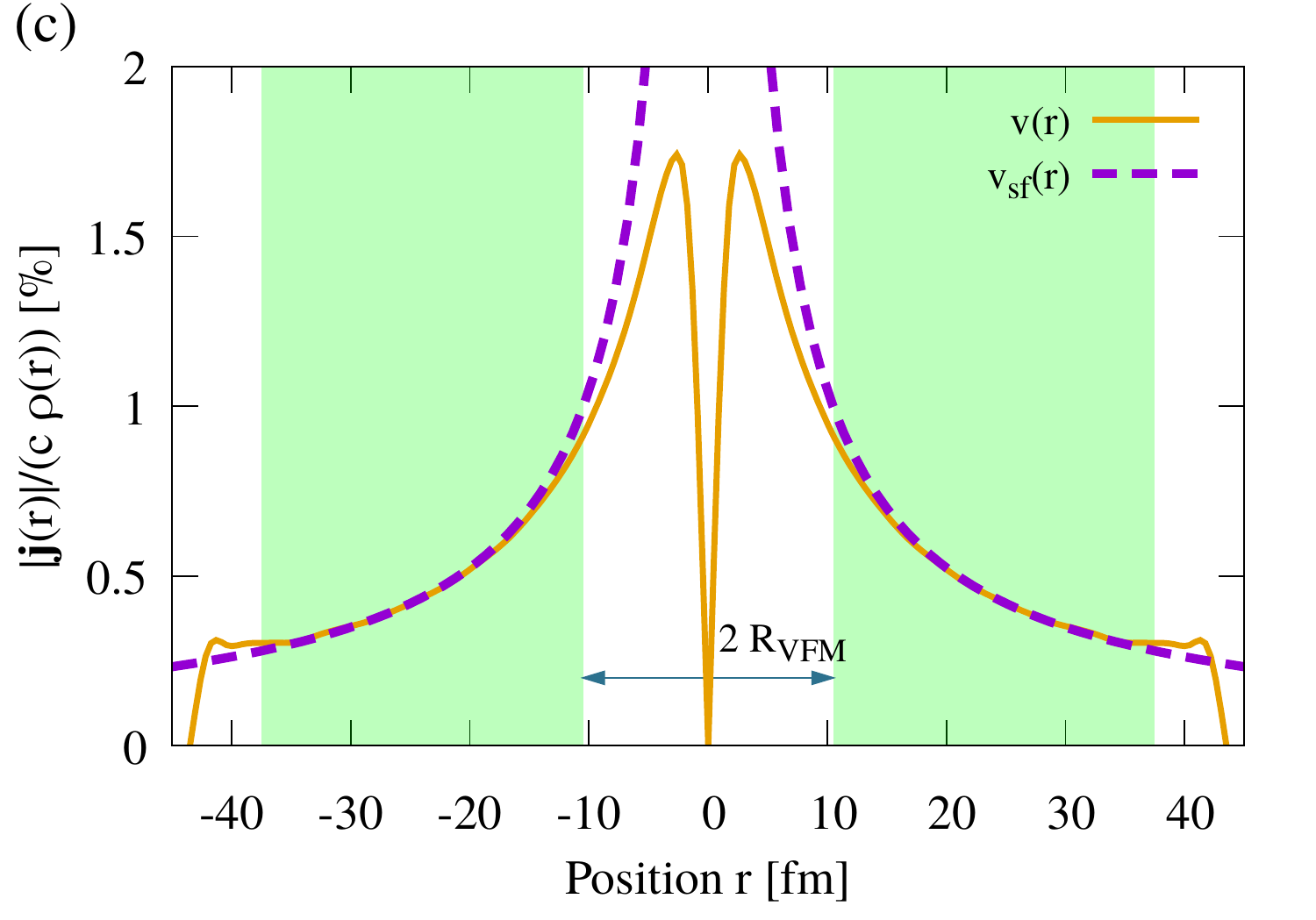}
 \caption{
(a) Norm of the velocity field $\pmb{v}(\pmb{r})$ (in units of $10^{-2} c$), and (b) of the current $\pmb{j}(\pmb{r})$ (in units of $10^{-5} c$ fm$^{-3}$) for the bulk neutron density $\rho=0.0059$ fm$^{-3}$ and the corresponding bulk pairing gap $\Delta_\infty=1.334$ MeV. 
 Three circles in the middle have radii: $k_F^{-1}, \xi, R_\mathrm{VFM}$, respectively. 
 The fourth circle near the boundary, with the radius $R_\mathrm{max}$, shows where the 
 the boundary effects becomes nonnegligible.
 (c) Cross section through the center of the vortex core. The orange solid curve denotes the velocity $\bm{v}(\bm{r})$ associated with the mass current given by Eq.~\eqref{eq:v}, while the purple dashed line depicts $\bm{v}_{\mathrm{sf}}(\bm{r})$.
 While the former has a finite value in the core, the latter diverges.
 The green area shows the region where the difference between the two velocities is smaller than $10\%$. The radius $R_\mathrm{VFM}$ is defined as the distance from the vortex core to the point where the curves start to match.
 \label{fig:velocities}}
\end{figure}

The superfluid velocity field $\bm{v}_{\mathrm{sf}}(\pmb{r})$ is related to the gradient of the phase 
$\phi(\pmb{r})$ of the condensate wave function, which is related to the pairing field $\Delta(\pmb{r})=|\Delta(\pmb{r})|e^{i\phi(\pmb{r})}$. The velocity thus acquires the following form:
\begin{equation}\label{eq:v_sf}
 \bm{v}_{\mathrm{sf}}(\bm{r})= \frac{\hbar}{2M}\bm{\nabla}\phi(\bm{r}) = \frac{\hbar}{2M} \frac{1}{r} \bm{\hat e_\phi}.
\end{equation}
Note that we have used 
the mass $2M$ of the Cooper pair which is twice the mass of the neutron. 
Another way to describe the flow is to define the velocity through its relation to the mass current~\cite{allard2020entrainment}:
\begin{equation}\label{eq:v}
 \bm{v}(\bm{r}) = \hbar \frac{\bm{j}(\bm{r})}{\rho(\bm{r})}.
\end{equation}
In case of absence of the normal component, it is identical to the superfluid velocity Eq.~\eqref{eq:v_sf}. 
Both definitions agree at large distances from the center for $r\gg \xi, r\gg k_F^{-1}$, where the impact of the vortex core is negligible, and $\bm v(\pmb{r}) \approx \bm v_{\mathrm{sf}}(\pmb{r})$,
but they differ if one approaches the core, see Fig.~\ref{fig:velocities}(c). Namely, at a certain distance the velocity of the superflow 
becomes comparable with Landau's critical velocity and the modulus of the pairing field starts 
to decrease as the center of the core is approached. 
This is precisely the transition point where the influence of the core is no longer 
negligible and affects the motion of neutrons. The identification of the transition point is crucial if one intends to properly define the VFM. 
The VFM ignores the complexity of the core structure, representing a vortex as a line
of vanishing thickness. It assumes that the velocity field decays with the distance 
from the vortex line as $\sim r^{-1}$. 
The finite thickness of the vortex line, still sits in the VFM as a parameter, which cannot be set to zero, as this would lead to divergences (the divergence is logarithmic in the so-called local induction approximation \cite{Barenghi}).
Therefore, we address below the question of determining the suitable vortex core size for the VFM. 

In Figs.~\ref{fig:velocities}(a) and \ref{fig:velocities}(b), we show the norm of the velocity $\pmb{v}$, and of the current $\pmb{j}$, respectively. Due to the geometry imposed, both quantities lay within a two-dimensional plane. To compare the values of the different length scales, we 
plot circles centered in the vortex core, having radii of: $k_{\mathrm{F}}^{-1}$, $\xi$, $R_{\mathrm{VFM}}$ (this is the radius for which the two definitions~(\ref{eq:v_sf}--\ref{eq:v}) coincide). Within the radius of $\xi$ the velocity doubles to roughly $2\%$ of the speed of light $c$, and then within the radius $k_{\mathrm{F}}^{-1}$ the velocity vanishes in the vortex core.

Fig.~\ref{fig:velocities}(c) presents the cross section of the velocity field through the vortex core. At large distances from the center, the expected behavior $\bm{v}_{\mathrm{sf}}(\bm{r})\sim r^{-1}$ is reproduced. However, deviations are seen at short distances. In particular, the velocity vanishes in the vortex core, instead of diverging like $\sim r^{-1}$. This reflects the  appearance of a normal component, even in the zero temperature limit. 
We compare both velocities in Fig.~\ref{fig:velocities}(c), and mark the area where the two velocities differ by less than $10\%$. The discrepancy far from the core at $x\approx\pm35$~fm comes purely from the boundary effects and the finite size of our simulation. However, we define the radius  $R_{\mathrm{VFM}}\approx 10.5$~fm from the vortex core, at which both definitions of velocity agree. For other densities, see Table~\ref{tab}.
Our calculations indicate that $R_{\mathrm{VFM}}$ is the relevant radius for the VFM, and is typically of order of a few coherence lengths $\xi$.

\begin{figure}
\centering
\includegraphics[width=0.99\linewidth]{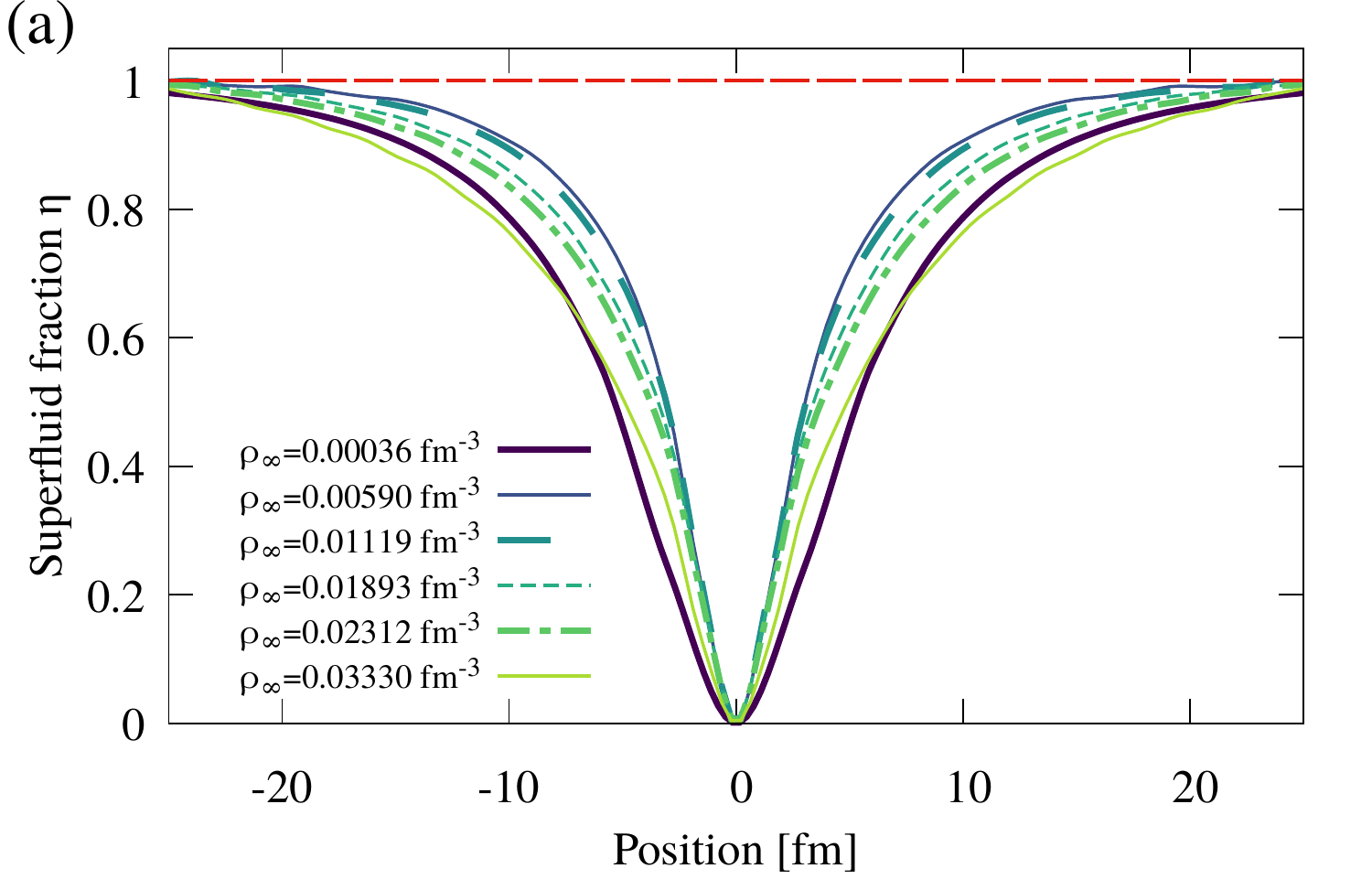}
\includegraphics[width=0.99\linewidth]{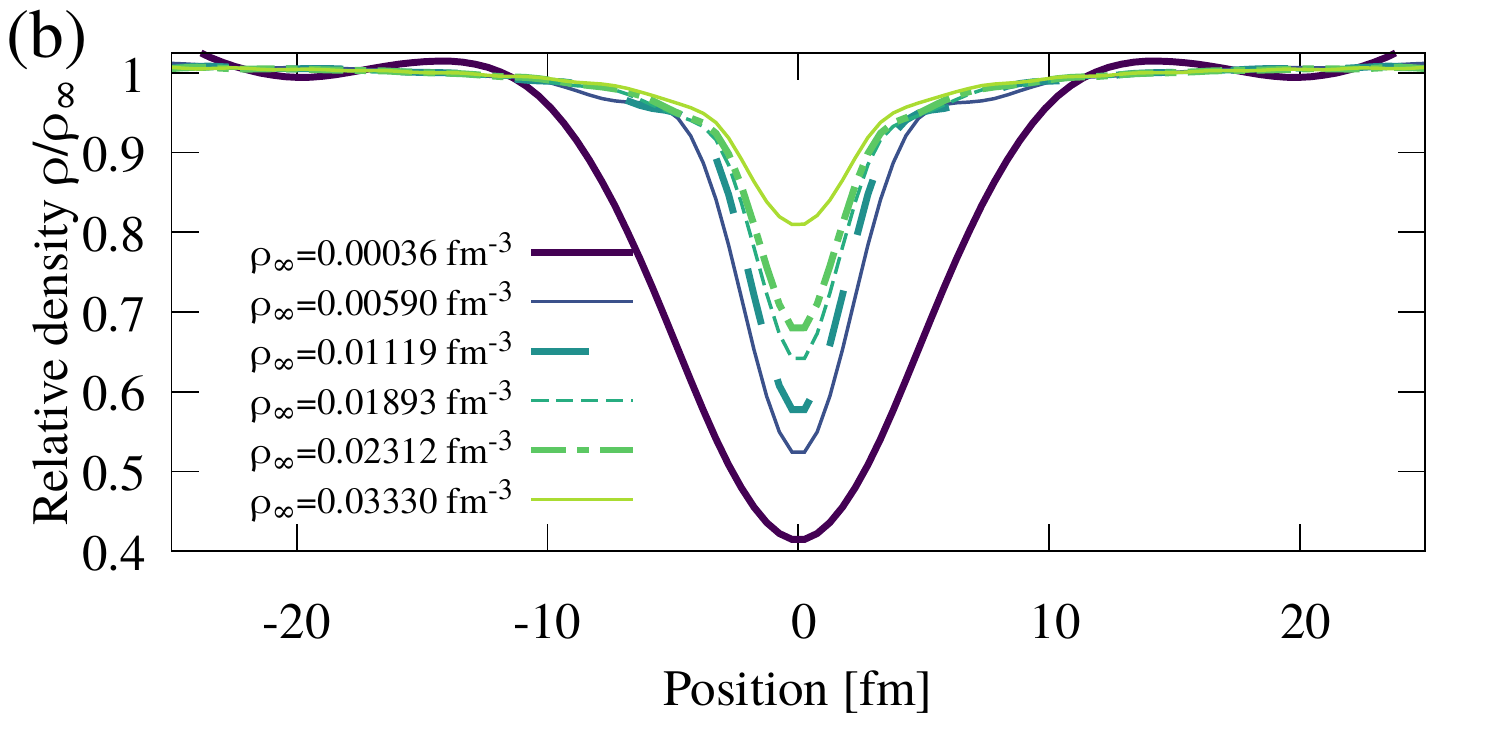}
\caption{
(a) Superfluid fraction $\eta=v(\pmb{r})/v_\mathrm{sf}(\pmb{r})$ for different densities $\rho$ in the zero temperature limit. Far from the vortex core it saturates to $1$, while it drops to zero in the core. 
(b) Density of the system relative to the bulk density $\rho_\infty$ far away from the vortex core. Note that the vertical axis does not start with 0, ie. the core is filled with matter. 
\label{fig:superfluid}}
\end{figure}

From the two definitions of the velocity field, one may introduce the superfluid
fraction as the amount of matter that locally takes part in the superflow.
Namely, the superfluid fraction $\eta=\rho_s/\rho$ can be estimated by the ratio 
$\eta=v(\pmb{r})/v_\mathrm{sf}(\pmb{r})$
We depict the value of $\eta$ across the cross-section of the vortex in Fig.~\ref{fig:superfluid}(a). Combining this with the density distribution, Fig.~\ref{fig:superfluid}(b), we clearly demonstrate that the vortices in neutron matter (and general in fermionic superfluids) carry some normal component in their core, even in the zero temperature limit. In the next section, we will demonstrate that this component induces modifications to the heat transport in the crust. 

\section{Heat capacity}\label{sec:heat_capacity}

The thermal electromagnetic emission from young neutron stars and from transiently
accreting neutron stars (see, e.g., Refs.~\cite{potekhin2015,wijnands2017} for 
recent reviews) does not directly reveal the properties of their dense core. This stems from the fact that the core 
remains thermally insulated by the hot crust, whose outermost layers form a heat-blanketing envelope. 
The (observable) time it takes to reach thermal equilibrium - typically
several decades - thus depends on the thermal properties of the crust, and in particular 
the ratio between the thermal conductivity (mainly due to electrons) and the specific heat.  The latter is a measure of the number of degrees of freedom, which in turn is determined by the composition and the structure of the crust. Although the internal constitution 
of the outer crust is fairly well known, that of the inner region is more uncertain and has been the subject of considerable theoretical efforts~\cite{blaschke2018}. The major part of the inner crust is expected 
to be made of a Coulomb lattice of spherical neutron-proton clusters immersed in a neutron 
superfluid and in a charge compensating relativistic electron gas. Due to the delicate interplay 
between nuclear and Coulomb interactions, the densest layers may consist of more exotic phases
referred to as nuclear ``pasta''. 

Since the specific heat coming from single quasiparticle excitations in a uniform 
superfluid is exponentially suppressed at temperatures much lower than the critical temperature (see, e.g., Ref.~\cite{pastore2015} for HFB calculations over the whole range of temperatures), the specific heat of superfluid neutron-star crusts is usually thought to be mainly determined by electrons and lattice vibrations (phonons)~\cite{potekhin2015} . The (volumetric) specific heat is thus approximately given by
\begin{equation}
\label{eq:specific-heat-crust}
C_V \approx 
\frac{1}{3} \frac{\varepsilon_{eF}^2}{(\hbar c)^3} T + \frac{2\pi^2}{15} \left(\frac{ T}{\hbar v}\right)^3  \, ,
\end{equation}
where $\varepsilon_{eF}$ is the electron Fermi energy, $v$ is the transverse 
phonon speed, and the temperature $T$ is assumed to be much lower than the Debye temperature. 
Longitudinal phonons have much higher velocities and therefore can be
neglected. 
However, pairing is a nonlocal phenomenon so that nuclear clusters may 
influence the quasiparticle excitations of the neutron superfluid, hence also the specific heat and cooling time \cite{pizzochero2002nuclear}.
Moreover, the inhomogeneity of neutron superfluid admits the presence of localized
in-gap states at the Fermi surface originating from the quasiparticle scattering on 
the pairing field \cite{magierski_andreev}. The effect of nonlocality has been addressed in the self-consistent HFB calculations in spherical Wigner-Seitz cells~\cite{monrozeau2007,fortin2010,pastore2015b} as well as fully three-dimensional band-structure calculations~\cite{chamel2009,chamel2010super}. More importantly, the lattice vibrations are influenced by the neutron superfluid e.g. originating from in-medium renormalization of the nuclear cluster masses~\cite{magierski_hydro1, magierski_hydro2}. The effective low-energy theory describing couplings between superfluid and solid matter has been formulated in Ref.~\cite{sanjay2011}. 
In particular, the transverse phonon speed is reduced and the corresponding specific heat, the second term in Eq.~(\ref{eq:specific-heat-crust}),  is increased~\cite{chamel2013}. Besides, longitudinal lattice vibrations are mixed with the low-energy collective excitations of the superfluid so that their combined contribution to the crustal specific heat may become comparable or even larger than that of electrons~\cite{chamel2013,chamel2016}. 
The superfluid excitations may be also coupled to nuclear-shape vibrations~\cite{inakura2017,inakura2019}. 
The latter become energetically favorable in the inner crust due to significant 
decrease of the nuclear surface tension which is expected to drop by order of 
magnitude as compared to normal nuclei.
At the bottom layers of the inner crust the situation become even more complex.
Increased susceptibility towards deformation of nuclear clusters results eventually
in the appearance of exotic structures involving very deformed nuclei (pasta phases)
\cite{magierski_shapes,casimir2,chamel2008}. 
Moreover due to shell effects~\cite{casimir1,casimir2,casimir3} it is expected that 
the long range order is lost. Consequently it is still not {\em a priori} known what are the 
degrees of freedom of nuclear matter that give the dominant contribution to specific heat.

In this paper, we focus on the influence of superfluid vortices on the neutron specific heat. 
The role of vortices is twofold. 
First, the neutron matter in the vortex core may be excited relatively easily as compared to 
the surrounding superfluid medium. Second, the superflow around the vortex modifies the quasiparticle
excitations and thus introduces corrections to the specific heat of a uniform superfluid.
Clearly, whether the vortex contribution is significant or not
depends on the density of topological excitations. Our aim, however, is to determine
the modification of the specific heat associated with a single vortex in a way
that will allow to make predictions for an arbitrary vortex surface density, provided 
it is much smaller than $1/\xi^2$, where $\xi$ is the coherence length. Before embarking 
on numerical calculations let us first
discuss the source of the increased specific heat associated with a vortex. 

Fermionic vortices admit the existence of so called Caroli-de Gennes-Matricon in-gap states~\cite{caroli1964,elgaroy2001superfluid}, which give rise to a finite density in the core. 
\begin{figure}[!ht]
\centering
\includegraphics[width=0.99\linewidth]{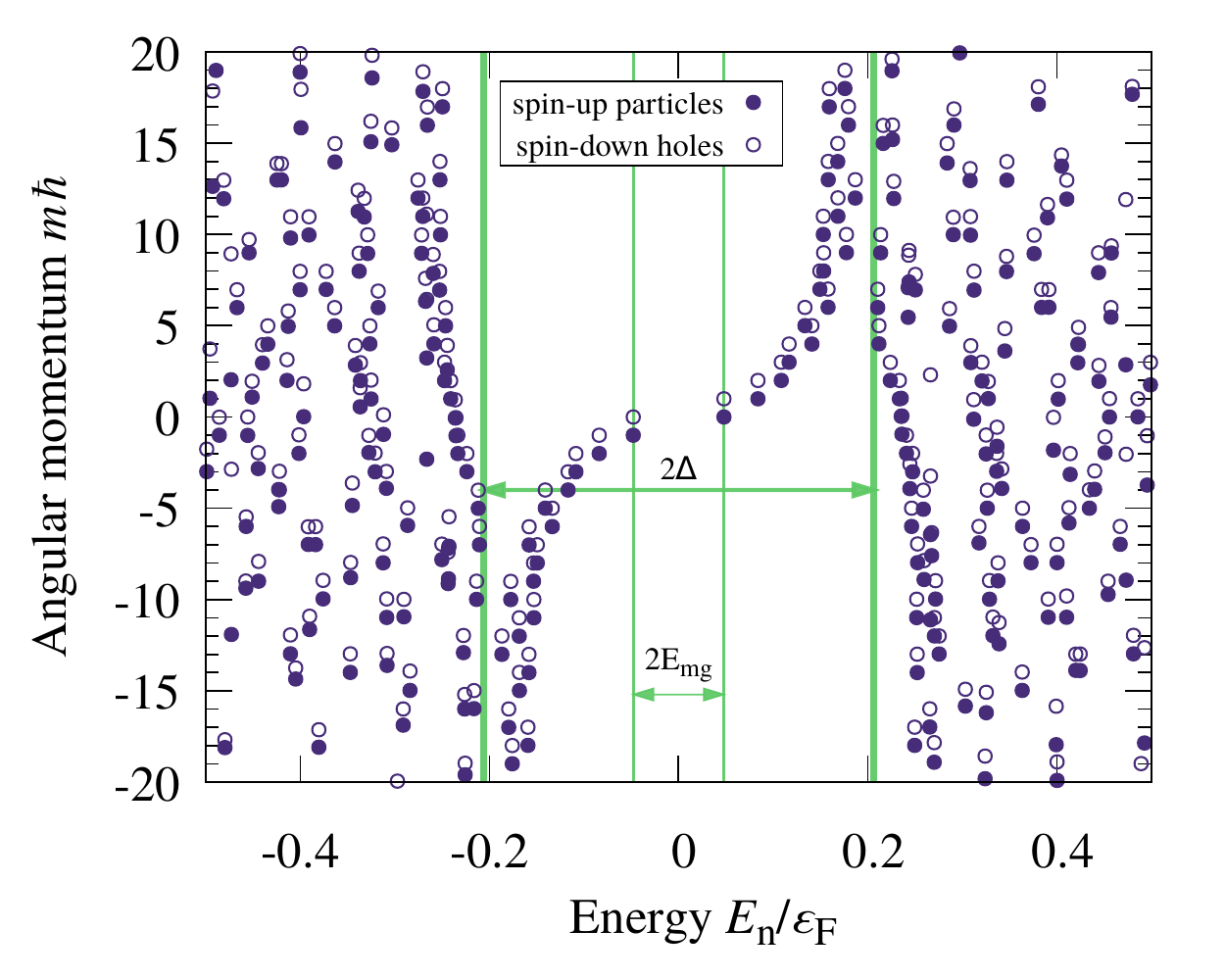}
\caption{
Distribution of angular momenta $m\hbar$ (along the vortex axis) of quasiparticle states of a given quasienergy $E_n$ (see Eq.~\eqref{eq:hfb}) in units of the Fermi energy $\eF=\hbar^2 k_F^2/(2 M)$. The finite density in the vortex core allows for the existence of in-gap states with energies within the pairing gap $\Delta$. The lowest energy state $E_{\mathrm{mg}}$ sets an important energy scale in the system. The data corresponds to the vortex simulation for the bulk density $\rho_\infty=0.00590 \mathrm{fm}^{-3}$.
\label{fig:andreev}}
\end{figure}
In Fig.~\ref{fig:andreev} we show a typical relation between the angular momentum per particle along the vortex axis $m\hbar$ ($m$ is the magnetic quantum number) of a given quasiparticle state, and its energy $E_n$. 
In a superconductor the pairing gap $\Delta$ sets the lowest energy scale. 
Here, due to the presence of the normal phase in the vortex core, the spectrum is modified and  subgap states of lower energy $| E_n | < |\Delta|$ are allowed. 
The energy of states associated with small enough angular momenta 
$\leq \hbar \kF \xi$ can be quite accurately reproduced within the semiclassical 
(Andreev) approximation, and follows with a good accuracy the formula~\cite{magierski_vortex} 
\begin{equation}\label{eq:subgap}
E_m=\frac{4}{3}\frac{|\Delta|^2}{\eF} m, 
\end{equation}
where $\eF=\hbar^2 \kF^2/(2 M)$ is the Fermi energy.
For larger angular momenta this band becomes flatter as a function of angular momentum
and their energies tend to $|\Delta|$ from below. Therefore, the contribution to
specific heat coming from large angular momentum states becomes smaller
and is exponentially suppressed for $T\ll|\Delta|$. As a consequence the
largest contribution to the specific heat comes from excitations localized inside
the vortex core. Note that due to the fact that these states have energies lying within 
the gap, they are formed by mixtures of particles and holes in almost equal proportions.
The number of these states increases with the coherence length as $\kF\xi$ 
reaching large values at the BCS side and vanishing in the vicinity of the unitary point
\cite{Sensarma2006,magierski_vortex}.

In a uniform superfluid, the existence of a pairing gap $\Delta$ suppresses exponentially the specific heat $C_V \propto e^{-|\Delta|/T}$. The presence of quantum vortices introduces states inside the gap. The lowest energy of subgap states so called minigap, which according to Eq.~(\ref{eq:subgap}) is given by 
\begin{equation}
\minigap=\frac{4}{3}\frac{|\Delta|^2}{\eF}\, , 
\end{equation}
thus becomes an important energy scale, see Fig.~\ref{fig:andreev}. 
It defines the characteristic temperature below which the specific heat will be exponentially suppressed. Moreover, the minigap also determines the minimum strength of the magnetic field that will effectively polarize the vortex core: for this to happen, a quasineutron excitation in the core
of energy $2\minigap$ is needed, leading to a critical magnetic field 
of the order of $B_{crit} \approx 2\minigap/(|\mu_{n}|)$, where $\mu_n\simeq -1.913 \mu_N$ is the neutron magnetic moment ($\mu_N$ being the nuclear magneton). Values of $\minigap$ and 
$B_{crit}$ extracted numerically for densities corresponding to neutron-star crusts are listed in Table~\ref{tab}. These values are comparable to those expected to be found in magnetars~\cite{kaspi2017}. Vortex-core polarization may thus occur in these strongly-magnetized neutron stars.  

Within the temperature range $\minigap\le T \le |\Delta|$, the 
specific heat is expected to
have a different form. In the far BCS limit where the interlevel spacing of core states
is negligibly small it should follow a linear dependence on $T$, the same as for a Fermi gas \cite{caroli1964}.
More precise estimates can be performed based on the density of Andreev states.
Namely, let us consider the regime in which the temperature is much smaller than the
critical temperature ($T \ll \Tc$) and at the same time is large compared to  the minigap, $T\gg \minigap$, so that excitations of subgap states are allowed. 
It follows that the heat capacity (per unit of length $L$) is linear in $T$ and is given by the relation:
\begin{equation}
 \frac{C_V}{L} = \frac{\pi}{8} k_F^2 \xi\frac{T}{|\Delta|} \propto T .
\end{equation}
In practice, however, as will be shown below, the finite interlevel spacing and the increase 
of the size of the vortex core with temperature lead to departure from this linear 
behavior. 

In order to evaluate the specific heat, we have applied the numerical setup
described in Sec.~\ref{num}. The energy of the vortex configuration as a function of temperature was computed, and next by taking derivative (using finite difference method) the corresponding specific heat $C_V$ was extracted.  
In the case of the tube with the vortex, the contribution to the energy per particle can be
expressed as the sum of three terms:
\begin{equation}\label{eq:E-contributions}
E = E_{\mathrm{vor}} + E_{\mathrm{core}} + E_{\mathrm{boundary}}.
\end{equation}
The first term corresponds to the uniform fluid with a velocity field generated
by the vortex, i.e., $\propto 1/r$. The second term is related to the core structure, filled with Caroli-de Gennes-Matricon states. 
Finally,
the last term is due to the boundary imposed in the calculations. It leads to the
fluctuation of neutron density close to the boundary. The wavelength of these spatial
oscillations is related to the inverse of Fermi momentum $1/\kF$, like in Friedel oscillations. The geometry of our setup is adjusted to ensure that oscillations due to boundary effects
and those due to the presence of the vortex core are spatially separated.
Therefore, the first two terms have physical origin, whereas the last one is associated with the numerical setup. The same kind of oscillations arise in the absence of a vortex: the density $\rho(\pmb{r})$ is 
almost uniform except near the boundary of the cylinder. 
To get rid of the spurious contribution to the specific heat coming from boundary effects,
we will thus consider the following difference:
\begin{equation}
\Delta C_V= C_V - C_V^{\mathrm{uniform}} = 
\frac{\partial  E}{\partial T} - 
\frac{\partial  E_{\mathrm{uniform}}}{\partial T},
\end{equation}
where $E_{\mathrm{uniform}}$ is the energy of the same system in the absence of a vortex.

\begin{figure}[!ht]
\centering
\includegraphics[width=0.99\linewidth]{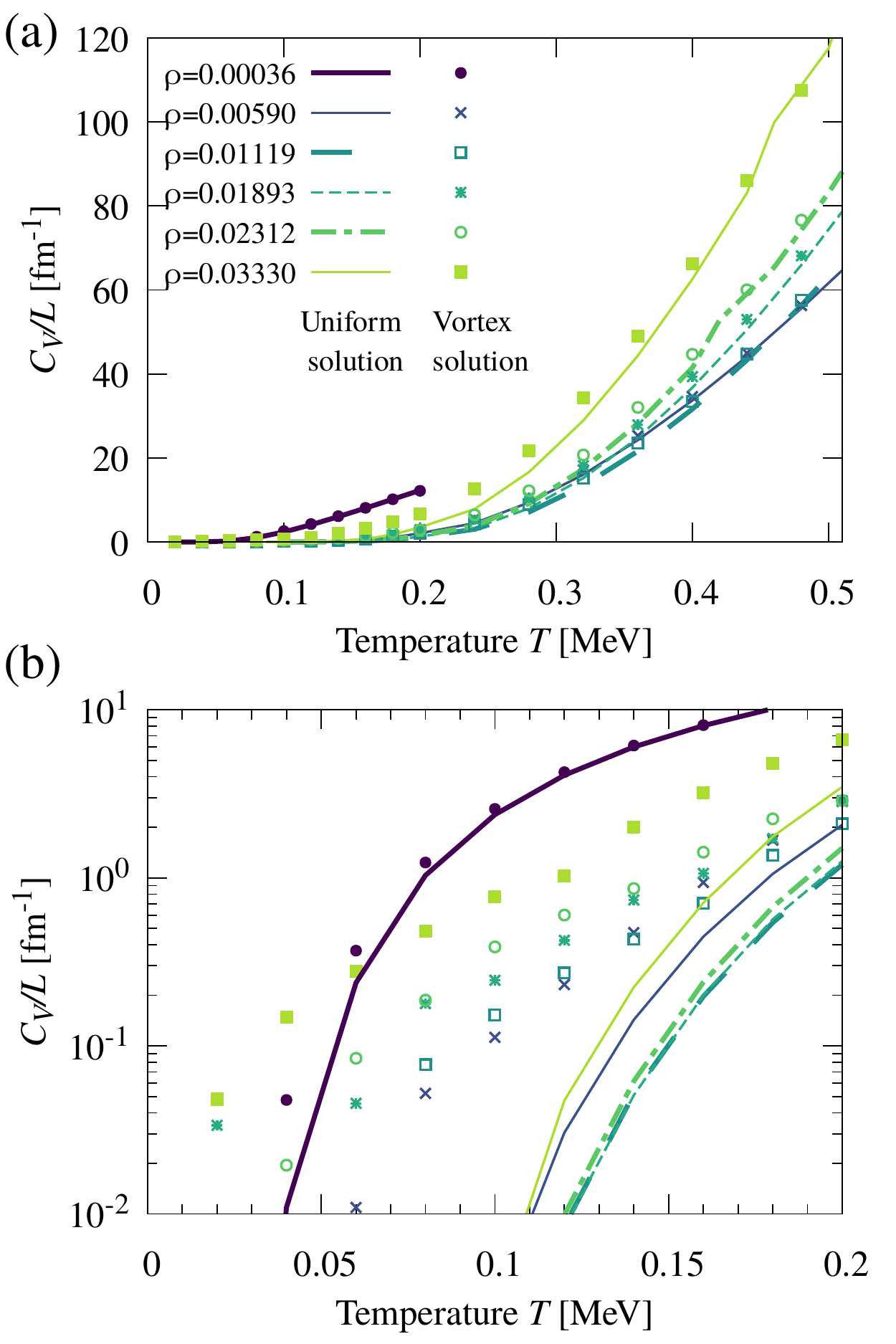}
\caption{
Specific heat per unit of vortex length $C_V/L$ in neutron matter for different densities $\rho$ as a function of the temperature $T$. The vortex and uniform solutions are denoted by points and lines, respectively. 
(a) The specific heat of the system with a vortex is systematically higher than that of  
the corresponding uniform system. Note that for the lowest density considered, the results are shown until
the pairing gap dropped by 25\% due to the temperature. For the remaining curves it corresponds to gaps equal approximately $\approx 1$~MeV.
(b) Specific heat in the vicinity of $T=0$ in logarithmic scale.
\label{fig:heat_capacity}}
\end{figure}

\begin{figure}[!ht]
\centering
\includegraphics[width=0.99\linewidth]{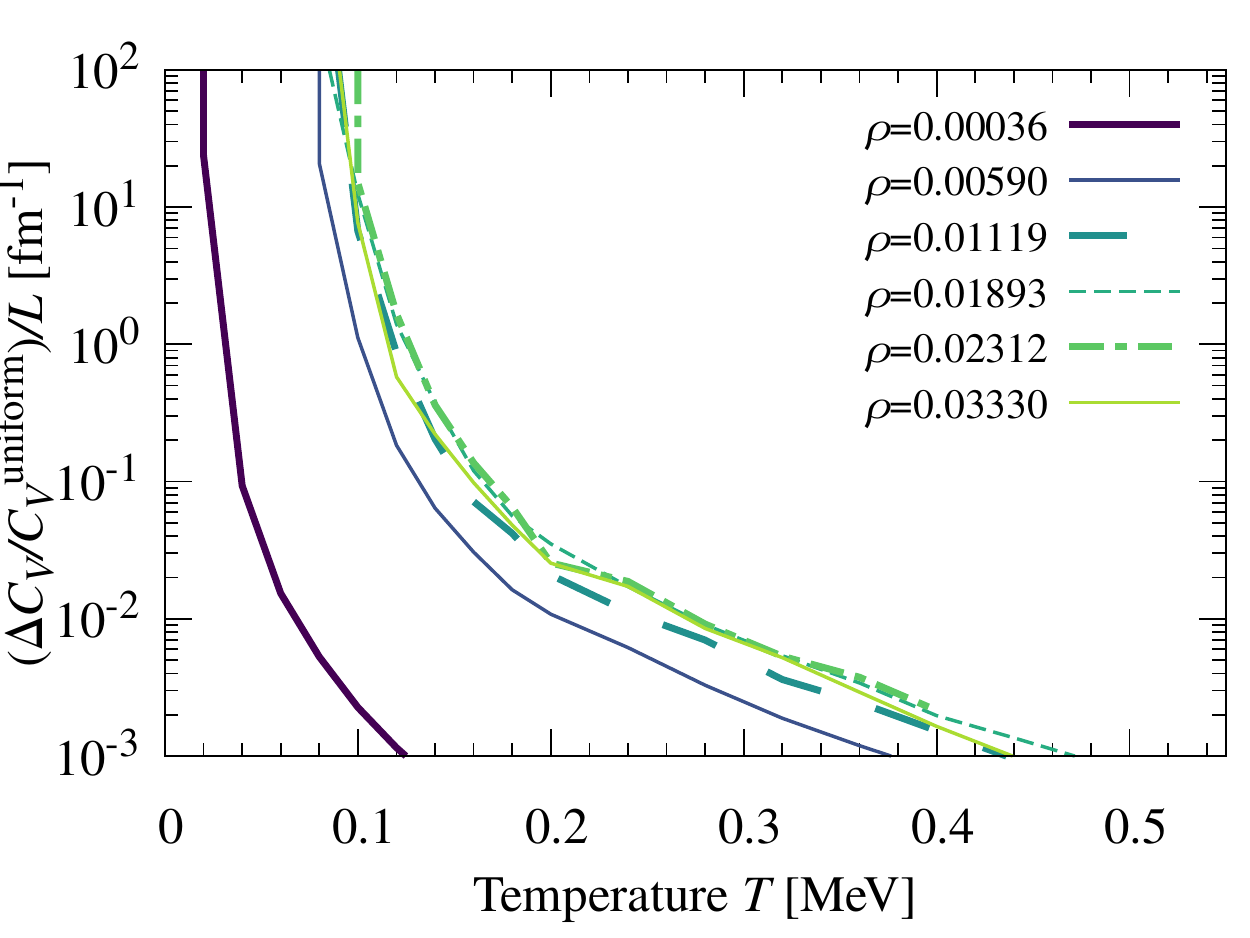}
\caption{
Ratio of the specific heat difference per unit length $\Delta C_V/L$ with and without a vortex to the specific heat of the uniform system $C_{V}^{\textrm{uniform}}$ for various neutron densities $\rho$ as a function of temperature $T$.  
The thick solid line stands out from the other due to much weaker pairing.
\label{fig:heat_capacity_ratio}}
\end{figure}

The specific heat per unit length of a vortex 
is shown in Fig.~\ref{fig:heat_capacity}(a) for a uniform system (lines), 
and for a vortex solutions (points) for different densities. 
The specific heat of the system with a vortex is found to be systematically higher in comparison to the uniform system over the whole range of temperatures. However, the deviations are the most pronounced at very low temperatures $T\ll \Tc$, as can be clearly seen in Fig.~\ref{fig:heat_capacity}(b). This expected result is due mostly to the presence of the vortex core inside of which superfluidity disappears, $|\Delta| \rightarrow 0$, as shown in Fig.~\ref{fig:core}. 
Clearly at temperatures $T$ approaching the critical temperature $\Tc$, the vortex ceases to exist and
therefore the specific heats with and without the vortex practically coincide. For the lowest density we considered, this corresponds to $T\lesssim 0.2$ MeV. For the sake of clarity, we have not displayed  results for temperatures such that the pairing gap has dropped by $25$~\% or more compared to its value at $T=0$. In the logarithmic scale adopted in Fig.~\ref{fig:heat_capacity}(b), the uniform solution varies as $-1/T$ (even when boundary effects are present), whereas the specific heat of a vortex solution is always considerably larger, and exhibits a different type 
of behaviors dictated by subgap states. To better assess the relative contribution of the vortex on the specific heat, we have plotted in Fig.~\ref{fig:heat_capacity_ratio} the ratio between $\Delta C_V/L$ and the specific heat $C_{V}^{\textrm{uniform}}$ of the corresponding uniform system. 
The results show clearly that the presence of a vortex may increase the specific heat of the neutron 
superfluid by several orders of magnitude at all densities provided the temperature is sufficiently low.

The result presented above are obtained for a cylinder of a certain radius, which in practice
is much smaller than the typical intervortex distance expected in neutron stars. However, in order to determine the specific heat of a system with vortices, one needs to take also into
account the correction coming from large distances from the vortex. 
This contribution is related to the presence of the superflow and arises from
the modification of the quasiparticle spectrum. Here, we present the procedure that
allows to infer a lower bound for the specific heat for a given vortex density. 
Namely, let us consider a large volume $S\times L$, where $S$ is the section area, containing a macroscopic piece of neutron-star crust corresponding to a neutron density $n$. Let us suppose that the area is threaded by $N_{\mathrm{vor}}$ vortices so that their surface density is 
$\sigma_\mathrm{vor}=N_{\mathrm{vor}}/S$. 
The (volumetric) specific heat difference between the configuration with and without vortices is proportional to $\sigma_\mathrm{vor}$:
\begin{equation}\label{cvmacro}
\frac{\Delta C^{\mathrm{macro}}_V}{S\cdot L} = \frac{N_{\mathrm{vor}}}{S}\left( \frac{\Delta C_V}{L} + \frac{\Delta C^{\mathrm{flow}}_V}{L}\right).
\end{equation}
The first term in the bracket arises from the quantum structure of each vortex core (as discussed above). It is calculated by considering a small cylinder with a typical radius of several tens of femtometers surrounding a single vortex. The second term represents the contribution 
coming from the vortex flow at large distances $r\gg\xi$. In  the absence of any vortex and ignoring mean-field effects, the quasiparticle energies for a uniformly moving superfluid with velocity $v_{s}$ are given by 
\begin{equation}
E_{\pm}(\pmb{k},\pmb{v_s}) = 
\frac{1}{2}\hbar\pmb{k}\cdot\pmb{v_s}\pm \sqrt{\left ( \frac{ \hbar^{2}k^{2}}{2M} - \mu \right)^{2}+|\Delta|^{2} }.
\end{equation}
In order to estimate the modification of the specific heat due to the vortex flow 
$v_{s}= \hbar/(2M r)$, we apply the Thomas-Fermi approximation, namely, 
we calculate the modification
of the specific heat locally, at the distance $r$ from the vortex and 
integrate the results over the cylindrical shell defined by the radii 
$R_{in}$ and $R_{out}$ (see Appendix~\ref{appendixB}): 
\begin{align} \label{cvflow}
\Delta C_{V}^{\mathrm{flow}} (R_{out}) &\approx 
\frac{1}{6}\frac{\hbar^2}{M |\Delta|}\frac{\mu}{|\Delta|} \frac{1}{R_{out}^{2}-R_{in}^{2}}
\ln\left ( \frac{R_{out}}{R_{in}} \right )
\times \nonumber \\ 
&\left [ \left (\frac{|\Delta|}{T}\right )^{2} - 4 \frac{|\Delta|}{T} + 2 \right ]C_{V}^{\textrm{uniform}},
\end{align}
where $C^{\mathrm{uniform}}_{V}$ is the specific heat of the uniform superfluid 
characterized by the pairing gap $\Delta$. In practice, $R_{in}$ and $R_{out}$ are 
of the order of the coherence length and the intervortex spacing respectively, therefore 
$R_{in}\ll R_{out}\sim \sqrt{S/N_\mathrm{vor}}/\pi$. 
Note that Eq.~\eqref{cvmacro} provides only a lower bound since the interactions between vortices, which we have neglected by treating each vortex independently, bring additional contributions to the specific heat.

From the equation \eqref{cvflow}, it is clear that the correction of the 
specific heat due to the vortex flow is positive. Assuming $T\ll |\Delta|$, this contribution 
is significant whenever 
\begin{align} \label{cond}
\frac{1}{6}\frac{\hbar^2}{M T}\frac{1}{R_{out}^{2}}\ln\left ( \frac{R_{out}}{R_{in}} \right ) \frac{\mu}{T} \approx 1, 
\end{align}
or equivalently, whenever the temperature is of the order of $T_v$ as given by 
\begin{align} \label{tempc}
\frac{T_v}{\mu}\approx \frac{1}{  \kF R_{out} } 
\sqrt{\frac{1}{3} \ln \left ( \frac{R_{out}}{\xi} \right )}, 
\end{align}
where we have approximated $\mu$ by the Fermi energy $\varepsilon_F$. 
However, it should be remarked that the contribution to the specific heat induced by the 
vortex flow is proportional to $C_{V}^{\textrm{uniform}}$, therefore it is exponentially 
suppressed in comparison to the specific heat contribution associated with the vortex 
core.

The procedure described above allows to estimate a lower bound for the specific heat associated 
with a given vortex density in neutron-star crusts, using the microscopic results 
presented in this paper and the formula (\ref{cvmacro}). The relative importance 
of the two terms in Eq. (\ref{cvmacro}) depends on the actual vortex density.
Given a typical surface vortex density of a neutron star $\sigma_\mathrm{vor}\approx 10^{-21} \mathrm{fm}^{-2} $, and the temperature of the crust $T/\Delta=0.1$, Eq.~\eqref{cvflow} yields that the flow related heat capacity $\Delta C_{V}^{\mathrm{flow}}$ is 16 orders of magnitude smaller than $C_{V}^{\textrm{uniform}}$. In view of the results shown in Fig.~\ref{fig:heat_capacity_ratio}, $\Delta C_{V}^{\mathrm{flow}}$ is therefore negligible compared to $\Delta C_V$. This stems from the very low surface density of vortices. However, the clustering of vortices due to the pinning might increase the contribution of vortex flow to the specific heat. 
Let us remark that this contribution may be substantial in physical systems studied in terrestrial laboratories, such as superfluid helium and ultracold fermionic condensates and for which the vortex core size is comparable with the intervortex spacing.

\section{Conclusions}\label{sec:conclusions}

In summary, we have carried out systematic fully self-consistent 3D HFB calculations of a quantum vortex in neutron matter for different temperatures and for various densities corresponding to different layers of the inner crust of a neutron star. We have adopted the BSk31 EDF from the latest series of the Brussels-Montreal family, which was specifically designed for such applications. 
We have determined the effective radius relevant for the VFM. 
Moreover, we have extracted the vortex core size 
as a function of temperature and have shown that it diverges at the critical temperature. We have shown that the superfluid fraction drops to zero inside the vortex core, thus indicating the presence of a normal phase. On the  contrary, the density remains finite due to the presence of so-called Caroli-de Gennes-Matricon in-gap states. We have determined their spectrum numerically and we have found that the magnetic field prevailing 
in magnetars may spin polarize the vortex core. 

We have also shown that the neutrons located in the vortex core are responsible for an additional contribution to the heat capacity. This effect considerably modifies the heat capacity at low  temperatures, i.e., temperatures comparable to the energy of minigap $E_\mathrm{mg}$.
At temperatures above the minigap but below the critical temperature the specific heat does not increase
linearly with $T$, as expected~\cite{caroli1964}. We attribute this effect to the dependence
of the core size on temperature and to the relatively low density of subgap states as compared to the BCS limit. 
Besides the contribution coming from vortex core state, we have obtained a lower  bound for the specific heat associated with the hydrodynamic flow induced by vortices. 
This contribution turns out to be negligibly small in neutron stars unless the surface density of vortices is sufficiently large.
On the other hand, the associated contribution may be substantial in terrestrial superfluids, where the density of vortices is large, even comparable with $1/\xi^2$.
Moreover, an additional increase of this term would stem from the fact that the flow farther than $R_{out}$ from the vortex cannot be neglected.

With the relatively large quasiparticle energy cutoff we have adopted here (thus ensuring energy conservation on long time scales), the wavefunctions we have computed could be  used  as initial data to study the time evolution of neutron vortices in neutron  stars, in particular vortex collisions and dynamical instabilities.

\begin{table}
 \caption{\label{tab} 
 Properties of neutron matter for different bulk densities $\rho_\infty$. The first three rows give characteristic length scales: the inverse Fermi momentum $k_F^{-1}$, the coherence length $\xi$, and the effective radius for the VFM $R_{\mathrm{VFM}}$. The following four rows give the energy scales: 
the bulk pairing gap $\Delta_\infty$ and corresponding critical temperature $\Tc$, the Fermi energy $\eF$, the chemical potential $\mu$, and the minigap $E_{\mathrm{mg}}$. The last row give the critical magnetic field $B_{\mathrm{crit}}$ needed to spin polarize the vortex core.
}
\begin{tabular}{r||c|c|c|c|c|c}
$\rho_\infty$  [fm$^{-3}$]& 0.00036 & 0.0059 & 0.0112 & 0.0189 & 0.0231 & 0.0333 \\ \hline
\hline
$k_F^{-1}$ [fm] & 4.52 & 1.79 & 1.45 & 1.21 & 1.14 & 1.01 \\ \hline
$\xi$     [fm] & 8.44 & 5.53 & 5.97 & 7.00 & 7.78 & 10.28 \\ \hline
$R_{\mathrm{VFM}}$ [fm] & 15.0 & 10.5 & 10.5 & 12.0 & 13.5 & 16.5 \\ \hline
$\Delta_\infty$ [MeV] & 0.35 & 1.33 & 1.53 & 1.55 & 1.50 &  1.28 \\ \hline
$\Tc$    [MeV] & 0.20 & 0.76 & 0.87 & 0.88 & 0.85 & 0.73 \\ \hline
$\eF$ [MeV] & 1.01 & 6.48 & 9.93 & 14.09 & 16.10 & 20.53 \\ \hline
$\mu$ [MeV]& 0.80 & 4.21 & 5.80 & 7.30 & 7.91 & 9.09 \\ \hline
$E_{\mathrm{mg}}$    [MeV] & 0.090 & 0.308 & 0.261 & 0.199 & 0.152 & 0.009 \\ \hline
$B_{\mathrm{crit}}$ [$10^{15}$ G] & 7.76 & 26.5 & 22.5 & 17.2 & 13.1 & 0.82 \\ \hline
\end{tabular}
\end{table}

\section*{Acknowledgements}
One of the authors (PM) would like to thank Centre for Computational Sciences at University of Tsukuba,
where part of this work has been done for hospitality. One of the authors (DP) acknowledges hospitality 
from Universit\'e Libre de Bruxelles.
This work was supported by the Polish National Science Center Grants No. 2017/27/B/ST2/02792 (PM and DP) and 2017/26/E/ST3/00428 (GW), by the Belgian Fonds de la Recherche Scientifique (NC) under Grant No. PDR T.004320. We acknowledge PRACE  for  awarding  us  access  to  resource  Piz  Daint based in Switzerland at Swiss National Supercomputing Centre  (CSCS),  decision  No.   2019215113.  The calculations were supported in part by PL-Grid Infrastructure and in part by Interdisciplinary Centre for Mathematical and Computational Modelling (ICM) of Warsaw University (grant No. GA76-13). We thank the PHAROS COST Action (CA16214) for partial support.

\appendix
\section{Self-consistent procedure} 
\label{appendix}
The procedure that we employ is built on the following single-particle Hamiltonian:
\begin{equation}\label{ap:eq:ham}
 \hat{h} = 
   -\pmb{\nabla} B\cdot \pmb{\nabla} +
 U_\rho + U_\tau + U_{\Delta\rho} + U_\pi
 - \frac{i}{2} \{\pmb{A}, \pmb{\nabla} \},
\end{equation}
where we omit the spin-orbit coupling term. In the crust of a neutron star, the density gradients are smaller than in finite nuclei therefore this term can be safely neglected, thus reducing the computational cost of our calculations. 

The mean-field potentials used in Eq.~\eqref{ap:eq:ham} are calculated using a standard procedure of performing variation over density $\rho$:
\begin{flalign}
U_\rho =
&  \frac{\partial\mathcal{E}_\rho(\rho)}{\partial\rho}, \\
U_\tau =
&  \frac{\partial\mathcal{E}_\tau(\rho,\tau,{\pmb{j}})}{\partial\rho}, \\
U_{\Delta\rho} =
&  \frac{\partial\mathcal{E}_{\Delta\rho}(\rho,\pmb{\nabla}\rho)}{\partial\rho}
 -\pmb{\nabla}\cdot \frac{\partial\mathcal{E}_{\Delta\rho}(\rho,\pmb{\nabla}\rho)}{\partial\left(\nabla\rho\right)},\\
U_\pi =
&  \frac{\partial\mathcal{E}_\pi(\rho,\pmb{\nabla}\rho,\nu)}{\partial\rho}
 -\pmb{\nabla}\cdot \frac{\partial\mathcal{E}_\pi(\rho,\pmb{\nabla}\rho,\nu)}{\partial\left(\nabla\rho\right)}.
\end{flalign}
The potential $U_\pi$ connected to the pairing energy is negligible and has been omitted in our calculations. It is worth noting that it contains gradients of anomalous density $\nu$ which has a kink in the vortex core. Therefore, it would make the numerical procedure less stable. 
The mean-field potential $B$ coming from the varation over the kinetic density $\tau$ has a straightforward connection to the effective mass:
\begin{equation}\label{eq:effective_mass}
 B 
 = \frac{\hbar^2}{2M^\oplus}
 = \frac{\delta\mathcal{E}}{\delta\tau}
 = \frac{\hbar^2}{2 M} 
 + \frac{\partial\mathcal{E}_\tau(\rho,\tau,{\pmb{j}})}{\partial\tau}.
\end{equation}
The consecutive components of the mean-field potential vector field ${\pmb{A}}$ is defined as the varation over three components of the current ${\pmb{j}}$:
\begin{equation}
 {\pmb{A}} = 
 \frac{\delta\mathcal{E}}{\delta{\pmb{j}}} = 
 \frac{\delta\mathcal{E}_\tau(\rho,\tau,{\pmb{j}})}{\delta{\pmb{j}}} = 
 \frac{\partial\mathcal{E}_\tau(\rho,\tau,{\pmb{j}})}{\partial{\pmb{j}}}.
\end{equation}

The latest Brussels-Montreal functionals introduce density dependencies in some couplings. Their full expression is given by Eqs.~(A.30) in the appendix of Ref.~\cite{chamel2009further}. Here we provide their form for spin-unpolarized matter:
\begin{flalign}
 \label{eq:a:crho}
 C^\rho(\rho)   =
 &- \frac{1}{4}  t_0 \left(x_0 - 1 \right)
 - \frac{1}{24} t_3 \left(x_3 - 1 \right) \rho^\alpha \\
 \label{eq:a:ctau}
 C^\tau(\rho)   = 
 &- \frac{1}{8}  t_1 \left(x_1 - 1 \right)
 + \frac{3}{8}  t_2 \left(x_2 + 1 \right) 
 \nonumber \\
 &- \frac{1}{8}  t_4 \left(x_4 - 1 \right) \rho^\beta
 + \frac{3}{8}  t_5 \left(x_5 + 1 \right) \rho^\gamma \\
 \label{eq:a:claplace}
 C^{\Delta\rho}(\rho)   = 
 &\phantom{-} \frac{3}{32}  t_1 \left(x_1 - 1 \right) 
 + \frac{3}{32}  t_2 \left(x_2 + 1 \right)
 \nonumber \\
  &\phantom{-} \frac{3}{32}  t_4 \left(x_4 - 1 \right) \rho^\beta
 + \frac{3}{32}  t_5 \left(x_5 + 1 \right) \rho^\gamma .
\end{flalign}

\section{Contribution to the specific heat generated by superflow} 
\label{appendixB}

Here, we estimate the contribution to the specific heat coming
from the hydrodynamic flow induced by a quantum vortex. 
The contribution will be calculated for sufficiently low temperatures 
$T$ such that $d\Delta(T)/dT \approx 0$.

Let us consider first a uniformly moving superfluid with velocity $v_{s}$.
The quasiparticle spectrum reads:
\begin{equation}
E_{\pm}(\pmb{k},\pmb{v_s}) = 
\hbar\pmb{k}\cdot\pmb{v_s}\pm \sqrt{\left ( \frac{ \hbar^{2}k^{2}}{2M} - 
\tilde{\mu} \right)^{2} + |\Delta|^{2} },
\end{equation}
where $\tilde{\mu} = \mu -\frac{1}{2}M v_{s}^{2}$ represents
the correction of the chemical potential due to the superflow.
We assume that $\hbar \kF v_{s}/|\Delta| \ll 1$, i.e. we consider 
velocities much smaller than the critical velocity. In that case 
$E_{+}(\pmb{k},\pmb{v_s}) > 0$.
One can calculate the specific heat of the moving superfluid 
starting from the derivative of the entropy: 
prescription:
\begin{align}
C_{V}
=-T\frac{ 2 V}{(2\pi)^{3}}
\int d^{3}k \frac{d}{dT} \ln\left [1 + \exp\left(\frac{-E_{+}(\pmb{k},\pmb{v_s})}{T} \right) \right ].
\end{align}
Recalling $d\Delta(T)/dT \approx 0$, we obtain
\begin{align}
C_{V}&\approx\frac{2}{T^{2}}
\frac{V}{(2\pi)^{2}}\frac{M}{\hbar^2}\sqrt{\frac{2M}{\hbar^2}}
\int_{-\mu}^{\infty} d\varepsilon \int_{-1}^{1} d(\cos\theta) \sqrt{\varepsilon+\tilde{\mu}}  \nonumber \\
&\times \frac{\exp(E(\varepsilon)/T)}{(1+E(\varepsilon)/T)^2}
\left ( \sqrt{2M(\varepsilon+\tilde{\mu})} v_{s} \cos\theta + \sqrt{|\Delta|^2 + \varepsilon^2} \right )^{2},
\end{align}
where $E(\varepsilon)=\sqrt{2M(\varepsilon+\tilde{\mu})} v_{s}\cos\theta + \sqrt{\varepsilon^{2}+|\Delta|^2}$ and $\varepsilon =\hbar^{2}k^{2}/(2M)-\tilde{\mu}$. 
In the limit $T/||\Delta|| \ll 1$, we can approximate $\exp(E(\varepsilon)/T)/(1+E(\varepsilon)/T)^2\approx \exp(-E(\varepsilon)/T)$ and arrive at:
\begin{align}
C_{V}&\approx
\int_{-\mu}^{\infty} d\varepsilon \int_{-1}^{1} d(\cos\theta) F(\varepsilon,\cos\theta)
\exp\left ( -\frac{\sqrt{\varepsilon^2+|\Delta|^2}}{T} \right ),
\end{align}
where 
\begin{align}
&F(\varepsilon,\cos\theta)=\frac{V}{(2\pi)^{2}}\frac{M}{\hbar^2}
\exp\left ( -\sqrt{\frac{\varepsilon+\tilde{\mu}}{2M}} \frac{M v_{s} \cos\theta}{T} \right ) \nonumber \\
& \times \sqrt{\frac{2M(\tilde{\mu} + \varepsilon)}{\hbar^2}} 
\left (  \sqrt{2M(\varepsilon+\tilde{\mu})} v_{s} \cos\theta + \sqrt{|\Delta|^2 + \varepsilon^2} \right )^{2}.
\end{align}
The function $F$ is relatively slowly varying with $\varepsilon$ as compared to $\exp\left ( -\sqrt{\varepsilon^2+|\Delta|^2}/T \right )$ and therefore to calculate the integral we substitute $F(0,\cos\theta)$.
Keeping terms up to $v_{s}^2$ one finally obtains:
\begin{align}
&C_{V}(v_{s}) \approx\frac{2}{T^2}V N(0) \sqrt{2\pi T|\Delta|} \exp\left ( -\frac{|\Delta|}{T} \right ) \nonumber \\
&\times \left [ |\Delta|^{2}+\frac{\mu Mv_{s}^{2}}{3} 
\left ( \left (\frac{|\Delta|}{T}\right )^{2}-4 \frac{|\Delta|}{T}+2-\frac{3}{4}\left (\frac{|\Delta|}{\mu}\right )^{2} \right )^{2} \right ] ,
\end{align}
where 
\begin{equation}
N(0)=\frac{1}{(2\pi)^{2}}\left ( \frac{2M}{\hbar^{2}} \right )^{3/2}\sqrt{\mu}
\end{equation}
is the density of states at the Fermi surface. 
The term proportional to $\left (|\Delta|/\mu \right )^{2}$ originates from the density
correction due to the modification of the chemical potential $\tilde{\mu}$ in the presence of superflow. 
However, this correction is at least an order of magnitude smaller than the other terms and therefore will be neglected. Clearly $C_{V}(v_{s}) = C_{V}(v_{s}=0) + \Delta C_{V}$, where the first term corresponds to the specific
heat of the static superfluid, whereas $\Delta C_{V}$ is the correction due to the superflow, and since $T/|\Delta| \ll 1$ it is positive.

To apply this formula to the flow induced by a vortex, we apply the local density approximation. 
We consider a single vortex inside a cylinder of radius $R_{out}$. In order to evaluate the correction
to the specific heat coming from the region between radii $R_{in}$ and $R_{out}$ we 
use the fact that
the magnitude of superfluid velocity behaves like 
$v_{s}(r)=\hbar/(2 M r)$, where $r$ is the distance from the vortex axis. The radius $R_{in}$ denotes
the distance from the core, where the fluctuations of density and pairing field become negligible. 
In this paper we take $R_{in}$ as the radius of the cylinder where the microscopic calculations have been performed. 
Subsequently, we divide the volume between $R_{in}$ and $R_{out}$ into infinitesimal concentric cylindrical shells (of length $L$) in which  the superfluid velocity is well defined. Integrating the contribution $dC_{V}( v_{s}(r) )$ of each shell from $r=R_{in}$ to $r=R_{out}$ yields 
\begin{align}
\Delta C_{V} \approx &\frac{1}{3} (2\pi)^{3/2} N(0)\sqrt{\frac{|\Delta|}{T}}\frac{\mu}{T}
\frac{\hbar^2}{2M} L \ln\left ( \frac{R_{out}}{R_{in}}\right )  \nonumber \\
&\times\left [ \left (\frac{|\Delta|}{T}\right )^{2}
-4 \frac{|\Delta|}{T} + 2 \right ] \exp\left ( -\frac{|\Delta|}{T} \right ) .
\end{align}
Equivalently, one may express this correction through the specific heat $C_{V}^{\mathrm{uniform}}$ of the uniform static superfluid, characterized by the density of states $N(0)$ 
and the pairing gap $\Delta$:
\begin{align}\label{AppBCV}
\Delta C_{V} (R_{out}) &\approx 
\frac{1}{6}\frac{\hbar^2}{M |\Delta|}\frac{\mu}{|\Delta|} \frac{1}{R_{out}^{2}-R_{in}^{2}}
\ln\left ( \frac{R_{out}}{R_{in}} \right )
 \nonumber \\ 
&\times\left [ \left (\frac{|\Delta|}{T}\right )^{2} - 4 \frac{|\Delta|}{T} + 2 \right ]C_{V}^{\mathrm{uniform}}.
\end{align}

With the above expression, we can estimate the minimal contribution to the specific heat coming from the hydrodynamic flow outside the vortex core for any given surface density of vortices $\sigma_\mathrm{vor}=N_{\mathrm{vor}}/S$.
Assuming that the vortices are independent and the flow contributes only up to the distance $R_{out}$, we get a lower bound for the heat capacity by multiplying Eq.~\eqref{AppBCV} by the number $N_\mathrm{vor}$ of vortices and substituting $R_{out}=1/\sqrt{\pi \sigma_\mathrm{vor}}$: 
\begin{equation}
\Delta C_{V}^{\mathrm{flow}} = N_{\mathrm{vor}} \Delta C_{V} \left (\sqrt{\frac{1}{\pi \sigma_\mathrm{vor}}} \right ).
\end{equation}

\bibliographystyle{apsrev4-1}
\bibliography{bibtexNS}

\end{document}